\documentclass[usenatbib]{mn2e}
\usepackage{epsfig}
\usepackage{astrojournals}
\usepackage{amssymb}
\usepackage{multirow}
\usepackage{multicol}
\usepackage[varg]{txfonts}
\usepackage{color}
\usepackage{url}

\usepackage{graphicx}
\usepackage{lscape}
\usepackage{amssymb}

\title[Pulse phase spectroscopy of V\,0332+53]{Transient X-ray pulsar V\,0332+53: pulse phase-resolved spectroscopy and the reflection model}

\author[Lutovinov et al.]{A.\,A.\,Lutovinov,$^{1}$\thanks{aal@iki.rssi.ru}
S.\,S.\,Tsygankov,$^{2,1}$ V.\,F.\,Suleimanov,$^{3,4}$ A.\,A.\,Mushtukov,$^{2,5,6}$
\newauthor V.\,Doroshenko,$^{3}$  D.\,I.\,Nagirner$^{6}$ and J.\,Poutanen$^{2}$ \\
$^{1}$Space Research Institute of the Russian Academy of Sciences, Profsoyuznaya Str. 84/32, Moscow
  117997, Russia \\
$^{2}$Tuorla observatory, Department of Physics and Astronomy, University of Turku, V\"ais\"al\"antie 20, FI-21500 Piikki\"o, Finland \\
$^{3}$Institut f\"ur Astronomie und Astrophysik, Kepler Center for Astro and Particle Physics, Universit\"at T\"ubingen, Sand 1,\\
 D-72076 T\"ubingen, Germany \\
$^{4}$Kazan (Volga region) Federal University, Kremlevskaja str., 18, Kazan 420008, Russia\\
$^{5}$Pulkovo Observatory of the Russian Academy of Sciences,
  Saint Petersburg 196140, Russia\\
$^{6}$Sobolev Astronomical Institute, Saint Petersburg State University,
  Saint Petersburg, Staryj Peterhof 198504, Russia\\
}

\begin{document}


\pagerange{\pageref{firstpage}--\pageref{lastpage}} \pubyear{2014}

\maketitle

\label{firstpage}

\begin{abstract} We present the results of the pulse phase- and
luminosity-resolved spectroscopy of the transient X-ray pulsar
V\,0332+53, performed for the first time in a wide luminosity range
$(1-40)\times 10^{37}$ erg s$^{-1}$ during a giant outburst
observed by the \textit{RXTE} observatory in Dec 2004 -- Feb 2005.
We characterize the spectra quantitatively and built the detailed
``three-dimensional'' picture of spectral variations with pulse phase
and throughout the outburst. We show that all spectral parameters
are strongly variable with the pulse phase, and the pattern of this
variability significantly changes with luminosity directly
reflecting the associated changes in the structure of emission
regions and their beam patterns. Obtained results are qualitatively
discussed in terms of the recently developed reflection model for
the formation of cyclotron lines in the spectra of X-ray pulsars.

\end{abstract}


\section{Introduction}

The observed emission of X-ray pulsars originates from one or two hot
regions, located near the neutron star magnetic poles where the
accreting matter is funneled to by the strong magnetic field of the
neutron star. The geometry of the emission regions depends mostly on
the accretion rate, magnetic field structure and strength, and can be
described as spots, mounds or columns \citep[see e.g.,][]{bs76,bur91,beck07,mush14}.
The geometry and the beaming properties of these hot regions shape
the observed X-ray spectrum, its pulse-phase and accretion-rate
dependences which can be studied observationally and used to uncover
intrinsic properties of the emission regions.

Significant progress in X-ray astronomy has been achieved in the
last decades with the \textit{Compton-GRO, RXTE, BeppoSAX, INTEGRAL} and
\textit{Swift} observatories, which allowed to study several dozens of X-ray
pulsars in details \citep[see, e.g., reviews
of][]{bil97,dalf00,cob02,fil05,lut09a,cab12}. A combination of
excellent timing capabilities of the \textit{RXTE} observatory with
its good spectral characteristics allowed also to perform a pulse
phase-resolved analysis for a number of X-ray pulsars \citep[see,
e.g.,][]{krey04,klo08,fur11}.

\begin{table*}
\noindent
\centering
\caption{Log of \textit{RXTE} observations (ordered by luminosity) used for the pulse phase-resolved
analysis of the X-ray pulsar V\,0332+53.}\label{tab1}
\centering
\vspace{1mm}
\small{
\begin{tabular}{c|l|c|c|l}
\hline\hline
\# & Observation & Date &  Luminosity $^{a}$ & Phase $^{b}$\\
& ID & MJD &  ($10^{37}$ erg s$^{-1}$) & \\
\hline
1 & 90089-11-04-04 & 53360.0 & $36.9\pm0.2$& m \\
2 & 90089-11-04-02G & 53358.6 & $35.4\pm0.2$& b \\
3 & 90089-11-04-00G & 53356.5 & $34.3\pm0.3$& b\\
4 & 90089-11-03-01G & 53354.7 & $32.8\pm0.3$& b\\
5 & 90089-11-03-03 & 53352.8 & $30.4\pm0.2$& b \\
6 & 90089-11-03-04 & 53353.7 & $28.5\pm1.1$& b \\
7 & 90427-01-03-01 & 53385.1 & $22.5\pm0.8$& d \\
8 & 90427-01-03-05 & 53387.0 & $20.2\pm0.1$& d \\
9 & 90427-01-03-06 & 53387.4 & $19.6\pm0.1$& d \\
10 & 90427-01-03-09 & 53388.4 & $18.6\pm0.4$& d \\
11 & 90427-01-03-12 & 53389.3 & $18.0\pm0.5$& d \\
12 & 90014-01-02-13 & 53390.4 & $ 17.3\pm0.1$& d \\
13 & 90089-11-02-03 & 53343.8 & $16.3\pm0.1$& b \\
14 & 90089-11-02-00 & 53342.8 & $15.4\pm0.1$& b \\
15 & 90014-01-03-020 & 53394.4 & $15.3\pm0.1$& d \\
16 & 90014-01-03-03 & 53395.3 & $ 14.3\pm0.1$& d\\
17 & Group 1: & & $12.7\pm0.2$& d \\
& 90014-01-04-00 & 53398.5 & &\\
& 90014-01-04-01 & 53399.6 & &\\
& 90014-01-04-02 & 53401.4 & &\\
18 & 90014-01-05-01 & 53407.6 &  $9.7\pm0.1$& d \\
19 & 90427-01-04-04 & 53413.8 &  $7.0\pm0.2$& d \\
20 & Group 2: & & $ 6.2\pm0.1$& d \\
& 90427-01-04-02 & 53414.3 & \\
& 90427-01-04-03 & 53413.8 & \\
& 90427-01-04-05 & 53414.8 & \\
21 & 90427-01-04-01 & 53416.6 & $ 5.0\pm0.1$& d \\
22 & Group 3: & &  $3.2\pm0.1$& d \\
& 90014-01-07-01 & 53419.4 & &\\
& 90014-01-07-02 & 53420.7 & &\\
& 90014-01-07-03 & 53420.7 & &\\
23 & 90014-01-07-00 & 53424.4 & $ 1.7\pm0.1$& d \\

\hline
\end{tabular}
\vspace{3mm}

\begin{tabular}{ll}
$^a$ & Luminosity in the 3--100 keV energy band assuming the source distance of $7$ kpc \\
$^b$ & Outburst phase: b -- brightening, d -- declining, m -- near the maximum \\
\end{tabular}}
\end{table*}

In this work we focus on the detailed pulse phase-resolved
spectroscopy for the X-ray pulsar V\,0332+53 in a wide dynamical
luminosity range $L_{37}\simeq 1-40$ (here $L_{37}=L_X/10^{37}$ erg s$^{-1}$,
assuming the distance to the source of 7 kpc, \citealt{neg99}). It
is important to emphasize, that such systematic analysis has not been
carried out before.

The X-ray pulsar V\,0332+53 was observed by the \textit{RXTE}
observatory during a powerful outburst in 2004--2005, that provided a
very good coverage of different outburst stages. An additional reason to
choose V\,0332+53 for the study of physical conditions and geometry
of the emission regions in X-ray pulsars and for an examination of a
recently developed reflection model \citep{pout13} is a presence of the
cyclotron absorption line (so-called cyclotron resonant scattering
feature, CRSF) and its higher harmonics in the source spectrum
\citep{mak90,pot05,krey05}. A study of the source pulse-averaged
spectra showed that the cyclotron line energy, as well as parameters
of the spectral continuum, significantly change with the source
luminosity \citep{tsy06,mov06,tsy10,nak10} along the outburst.
This behaviour was interpreted as a result of changes in the height
of the accretion column where the cyclotron absorption line was assumed
to originate \citep[see, e.g.,][and references therein]{mih95,schon07,nish08,nish14,Becker2012}.
However, alternative scenario has recently been proposed by \cite{pout13},
where the line forms in the illuminated atmosphere of the neutron star.
In our paper, we focus on the latter model when discussing observed spectral
variations.

According to the model calculations the cyclotron line parameters
should strongly depend on the viewing angle, pulse phase, bulk
velocity of the falling plasma, etc.
\citep{ise98,arahar99,arahar00,serb00,schon07}. Parameters of the
continuum should also depend on the viewing angle due to the
significant angular dependence of the continuum
opacity in the strong magnetic field near the neutron star surface
\citep[see, e.g.,][]{kam82,kam83}. Thus, observations of V\,0332+53
give us an unique possibility to build for the first time a detailed
``three-dimensional'' picture (parameter {\rm vs} pulse phase
{\rm vs} luminosity) of the source spectrum behaviour, and ultimately
uncover the physical and geometrical picture of the accretion onto
a magnetized neutron star.

Note, that first results on the pulse phase-resolved spectroscopy for
V\,0332+53 were obtained by \cite{pot05}, however, available
observations did not contain the \textit{PCA} data with the required
energy and time resolution, so the analysis was limited to the
\textit{HEXTE} data and, therefore, it was not trivial to constrain
the parameters of the broadband spectrum. As a result, no significant
variations of spectral parameters with the pulse phase had been
reported. Later, some preliminary results (using only two observations
of V\,0332+53) on the phase and luminosity dependence
of the continuum and CRSF parameters were published
by \citet{lut09b} and the results on variability of the iron
emission line with the pulse phase and luminosity by \citet{tsy09}.
Also no attempts to interpret the obtained results have been made.

\section{Observations, spectral model and data analysis}

\begin{figure*}
\hbox{
\includegraphics[width=\columnwidth,bb=43 150 584 710, clip]{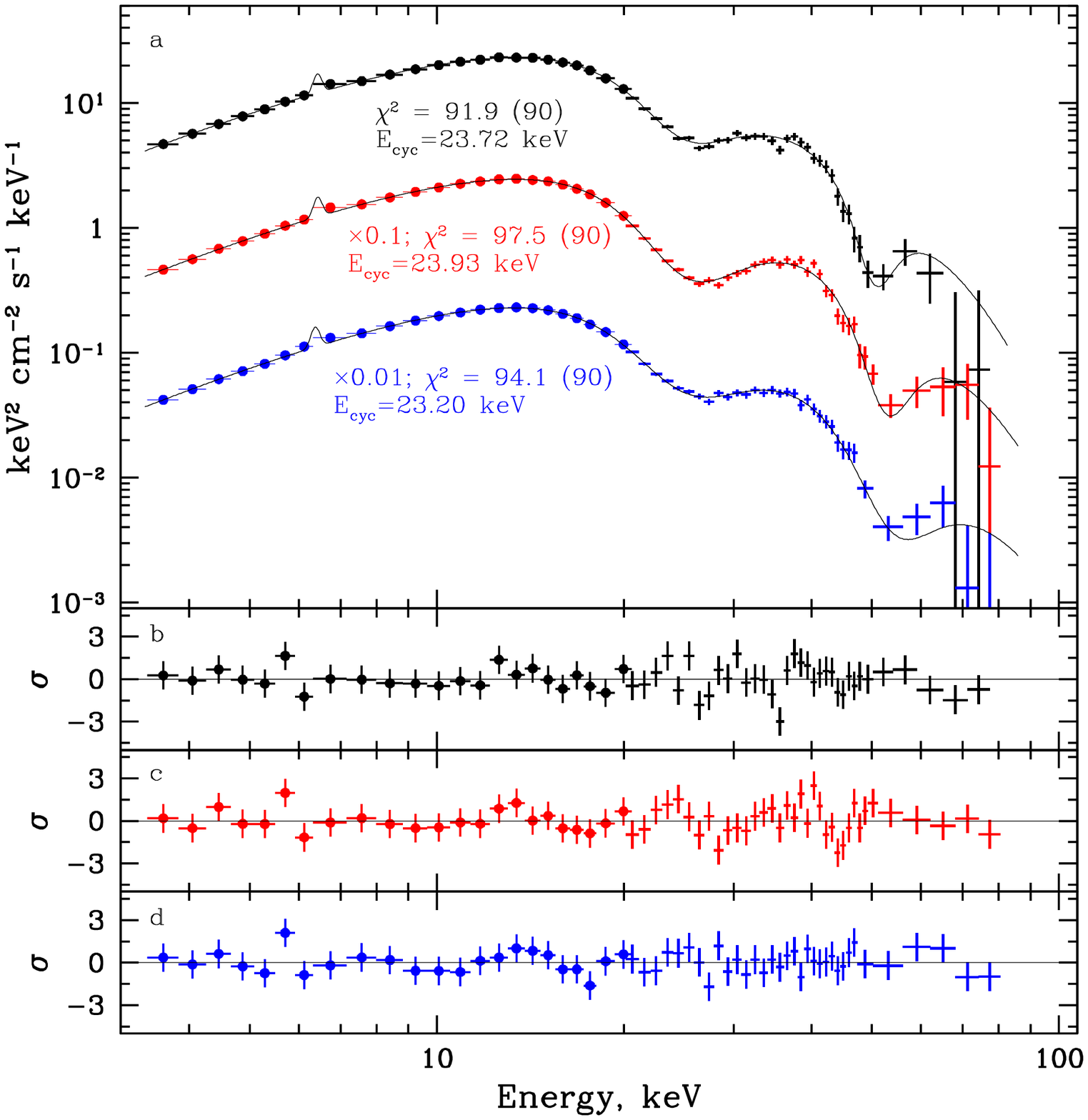}
\hspace{2mm}\includegraphics[width=\columnwidth,bb=43 150 584 710, clip]{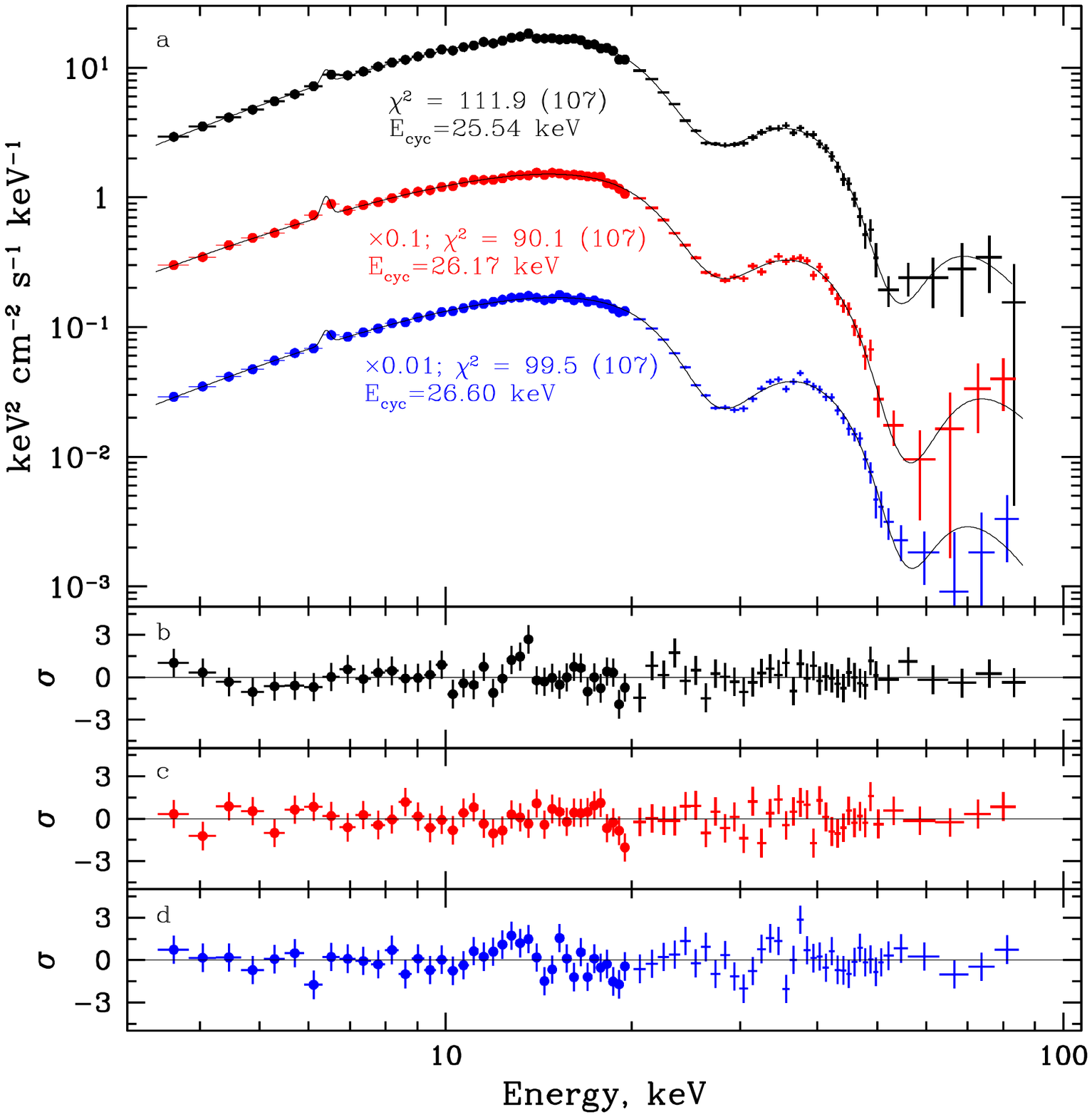}
}
\caption{
Representative examples of the pulse phase-resolved spectra of V\,0332+53
for two different luminosities $L_{37}=36.9$ (obs. 1, left)  and $20.2$ (obs. 8, right).
Black symbols correspond to the spectra at phases
0.5625--0.625 (left) and 0.75--0.8125 (right), red symbols correspond to the
spectra at phases 0.9375--1.0 (left) and 0.5--0.5625 (right), the blue symbols
are for the phases 0.5--0.5625 (left) and 0.375--0.4325 (right).
Values of $\chi^2$ and the cyclotron line energy are also shown for reference.
Residuals from the best-fitting models (summarised in Table\,\ref{tab2}) are shown in lower panels.
}
\label{ex_spec}
\end{figure*}

\begin{table*}
\noindent
\centering
\caption{Best-fitting parameters for three pulse phase-resolved spectra
for two different luminosities of V\,0332+53.}\label{tab2}
\centering
\vspace{1mm}
\small{
\begin{tabular}{l|c|c|c|c|c|c|c|c|r}
\hline\hline
Phase  & $\Gamma$ & $E_{\rm cut}$ & $E_0$ & $\sigma_0$ & $\tau_0$ & $E_1$ & $\sigma_1$ & $\tau_1$ & $\chi^2$ (d.o.f.) \\
    &          & keV            & keV      & keV           &            & keV      & keV           &            &                   \\
\hline
\multicolumn{10}{c}{Obs.1, $L_{37}=36.9$} \\ [0.5mm]
0.5625--0.625 & $0.35\pm0.07$ & $7.3^{+0.7}_{-0.5}$ & $23.72\pm0.15$ & $6.57^{+0.72}_{-0.16}$ & $1.47\pm0.08$ & $51.6^{+2.6}_{-1.5}$ & $8.2\pm3.8$ & $1.83^{+0.64}_{-0.34}$ & 91.9 (90)\\
0.9375--1.0 & $0.42\pm0.05$ & $7.6^{+0.5}_{-0.4}$ & $23.93\pm0.11$ & $6.68^{+0.47}_{-0.18}$ & $1.93\pm0.08$ & $51.6^{+1.2}_{-0.8}$ & $5.9^{+2.4}_{-1.8}$ & $2.56^{+0.92}_{-0.45}$ & 97.5 (90)\\
0.5--0.5625 & $0.18^{+0.08}_{-0.04}$ & $8.95^{+1.24}_{-0.44}$ & $23.20^{+0.18}_{-0.08}$ & $7.15^{+0.85}_{-0.38}$ & $1.38\pm0.08$ & $50.7^{+1.8}_{-1.1}$ & $10.5\pm2.8$ & $2.62^{+0.68}_{-0.30}$ & 94.1 (90)\\
\multicolumn{10}{c}{Obs.8, $L_{37}=20.2$} \\
0.75--0.8125 & $0.34\pm0.07$ & $8.5^{+0.5}_{-0.7}$ & $25.54^{+0.15}_{-0.17}$ & $6.00^{+0.24}_{-0.55}$ & $1.81\pm0.07$ & $52.2^{+1.2}_{-0.9}$ & $9.0^{+2.1}_{-1.3}$ & $2.37\pm0.32$ & 111.9 (107)\\
0.5--0.5625 & $0.28\pm0.08$ & $8.6^{+1.0}_{-0.9}$ & $26.17\pm0.27$ & $7.15^{+0.58}_{-0.92}$ & $1.88\pm0.13$ & $53.4^{+1.4}_{-1.1}$ & $5.5^{+2.7}_{-2.6}$ & $4.0^{+5.4}_{-1.2}$ & 90.1 (107)\\
0.375--0.4325 & $0.38\pm0.08$ & $7.7\pm0.7$ & $26.60^{+0.22}_{-0.19}$ & $5.41^{+0.53}_{-0.79}$ & $1.76\pm0.09$ & $54.4^{+1.7}_{-0.9}$ & $6.4\pm2.5$ & $2.38^{+0.83}_{-0.47}$ & 99.5 (107)\\
\hline
\end{tabular}
}
\end{table*}

\begin{figure*}
\hbox{
\includegraphics[width=\columnwidth,bb=45 155 550 690, clip]{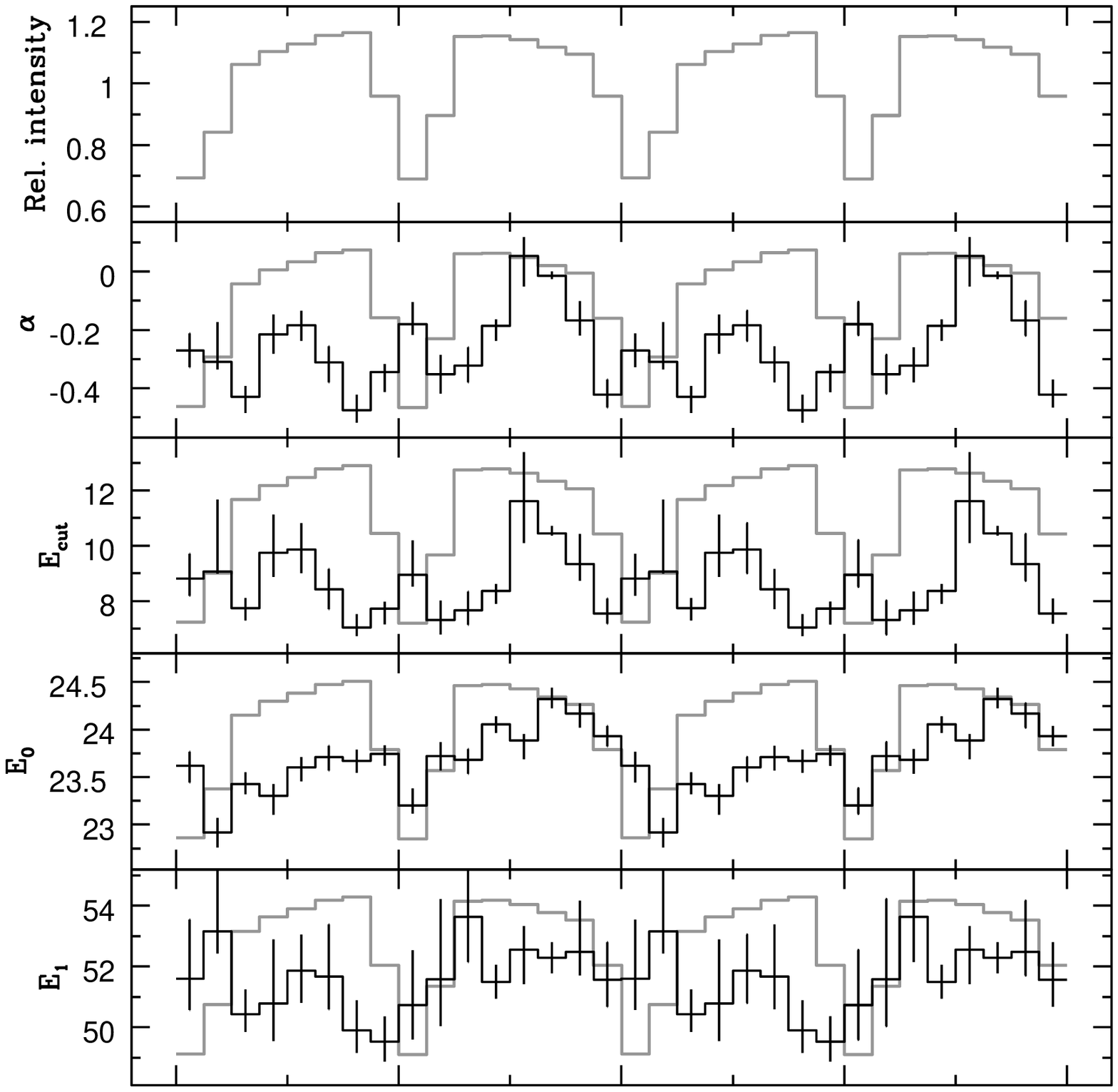}
\includegraphics[width=\columnwidth,bb=45 155 550 690, clip]{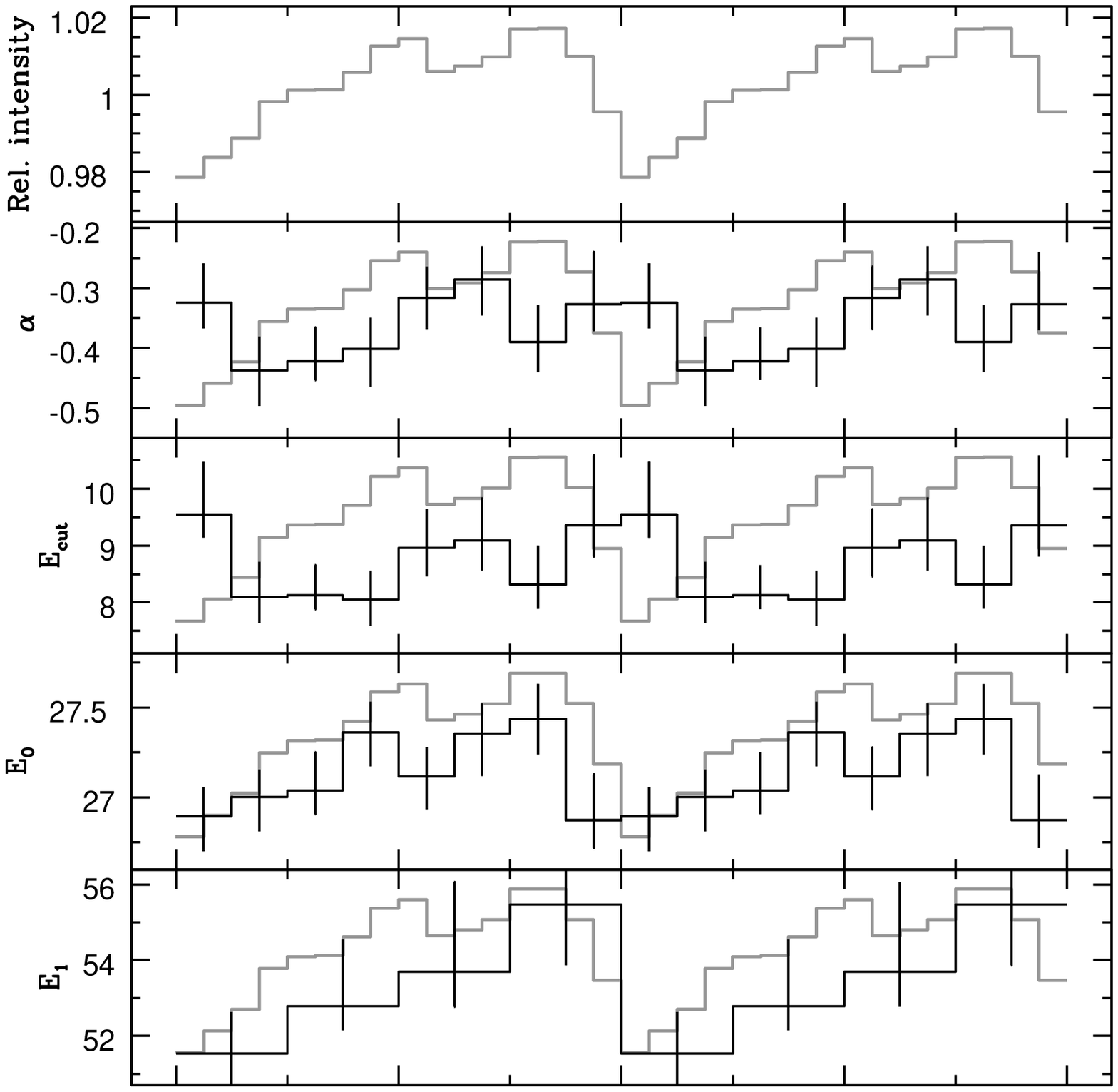}
}
\vspace{-9.3mm}\hbox{
\includegraphics[width=\columnwidth,bb=45 155 550 690, clip]{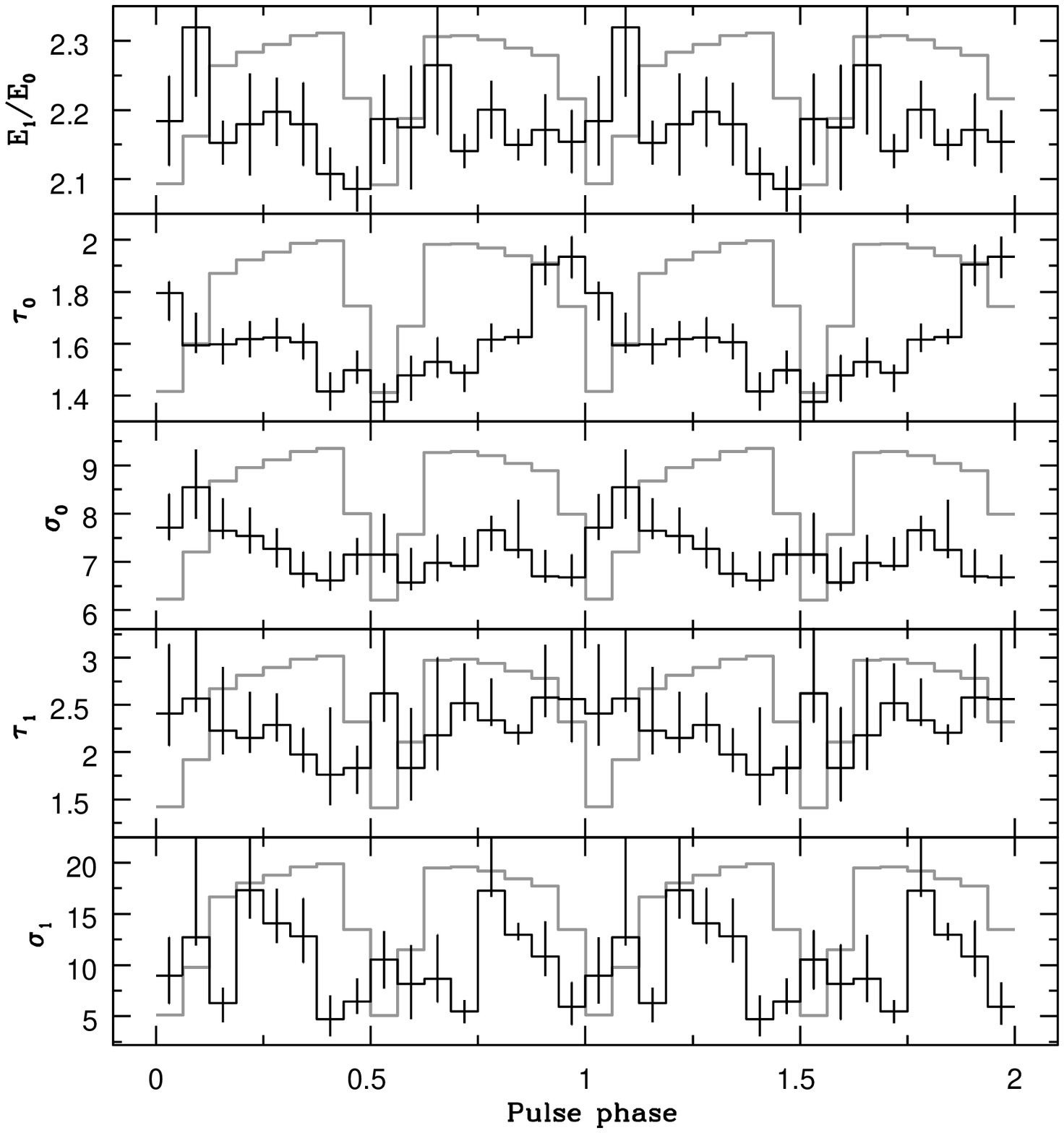}
\includegraphics[width=\columnwidth,bb=45 155 550 690, clip]{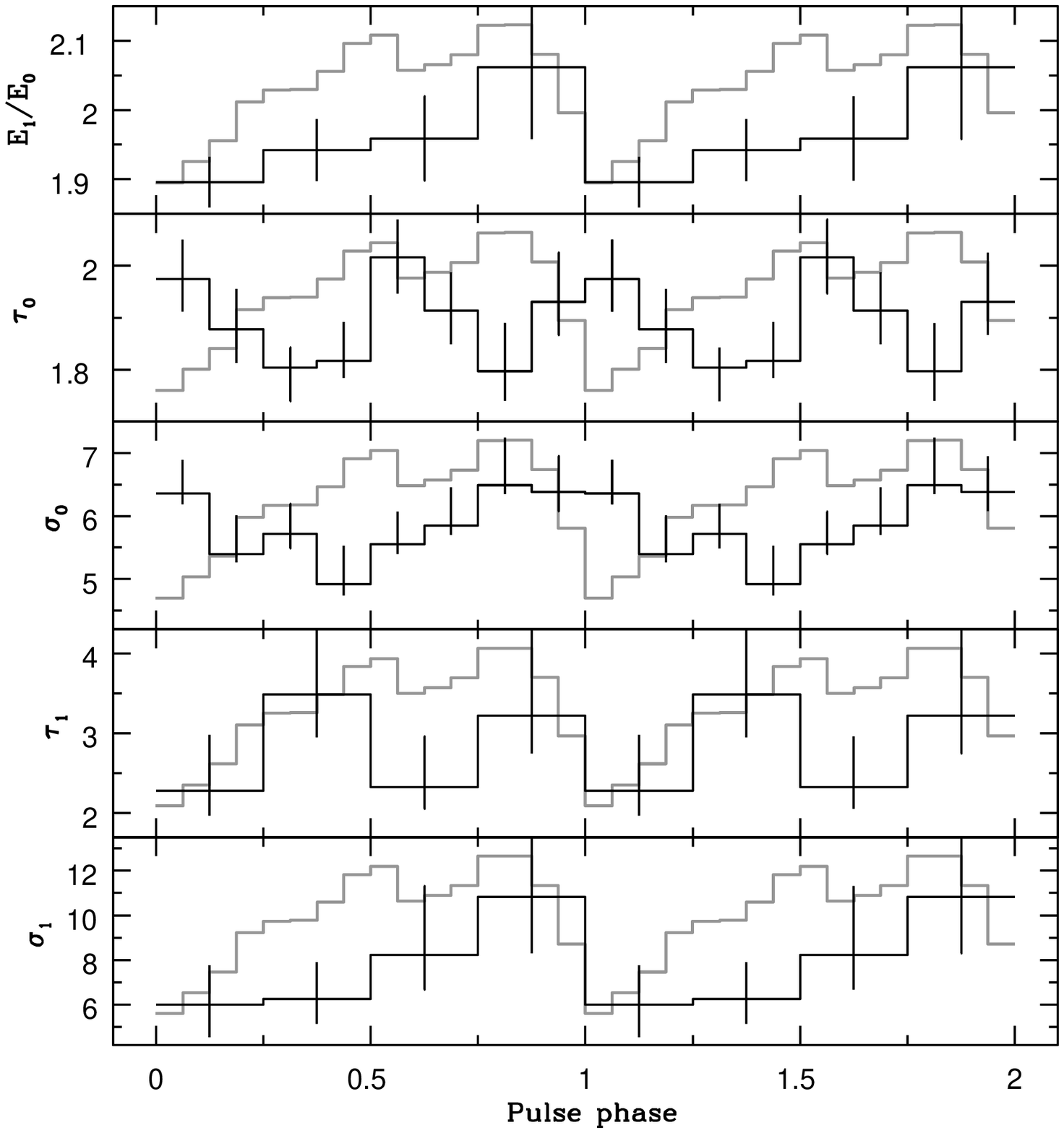}
}
\caption{Dependence of spectral parameters of V\,0332+53 on the pulse phase for
luminosities of $L_{37}=36.9$ (obs. 1, left) and
$L_{37}=12.7$ (obs. 17, right). Corresponding pulse profiles are
shown by black (upper panels) and grey (other panels) lines. Uncertainties correspond
to the $1\sigma$ confidence level. Two periods are shown for clarity.
}\label{pulphas_ex}
\end{figure*}

As already mentioned, V\,0332+53 has been observed with the
\textit{RXTE} observatory during the so-called giant outburst, which
started in the end of 2004 \citep{swa04} and lasted for about two
months. In the current work we use the same data set as in
\cite{tsy10} apart from a number of observations (ObsIDs.
90014-01-01-XX, 90089-11-05-XX, 90089-22-01-XX, 90427-01-01-XX,
90427-01-02-XX), performed with time resolution insufficient for the
pulse-phase resolved analysis (typically 16 s, while the source pulse
period is about of 4.35 s). The used data set covers both brightening
and declining phases of the outburst more or less uniformly. There is
only a small gap between luminosities of $L_{37}=22.5$ and
$L_{37}=28.5$ due to the absence of observations
of the required timing quality. To improve the statistic for
observations at lower luminosities or with small exposures we grouped
closely spaced observations with approximately the same averaged
source intensity into three groups with the averaged luminosities of
$L_{37} \simeq 3.2$, $6.2$ and $12.7$. The list of analysed observations with
corresponding averaged luminosities in the 3--100 keV energy range is
presented in Table \ref{tab1}. These luminosity values may be
regarded as close to the bolometric ones under the assumption that the
bulk of the energy is released in the X-rays.

To extract pulse phase-resolved spectra from the PCA and
HEXTE instruments onboard \textit{RXTE} we used standard programs of the
{\sc ftools/lheasoft} v. 6.14 package and methods described in the
\textit{RXTE}
cookbook\footnote{http://heasarc.nasa.gov/docs/xte/recipes/cook\_book.
 html}. Depending on the spectrum quality photons were folded
into 16, 8 or 4 phase bins, using individually determined (per
observation) and barycentrically corrected pulse periods.

The obtained spectra were approximated by a powerlaw model with an
exponential cutoff  ({\sc cutoffpl} in the {\sc xspec} package)
\begin{equation}
AE^{-\Gamma}\exp{(-E/E_{\rm cut})},
\end{equation}
where $\Gamma$ is the photon index, $E_{\rm cut}$ is the
cutoff energy, $A$ is a normalization. The spectrum was modified by one or several
cyclotron absorption lines in the form
\begin{equation}
\exp\left(\frac{-\tau_N(E/E_N)^2\sigma_N^2}{(E-E_N)^2+\sigma_N^2}\right),
\end{equation}
where $E_N$, $\sigma_N$ and $\tau_N$ are the energy,
width and depth of the cyclotron line ($N=0$) and harmonics ($N>0$),
respectively ({\sc cyclabs} in the {\sc xspec}
package; \citealt{mih90}). The same model was used earlier for the analysis of
pulse-phase averaged spectra and reasons for the choice of the continuum and line models
were discussed by \citet{tsy06,tsy10}. Note, that inclusion of the
photoelectric absorption in the model  does not affect the results of fitting
in the 3--100 keV energy range and thus it was not taken into account.

The first harmonic of the cyclotron absorption line was not
confidently observed and its parameters were not well constrained in
all data sets. Therefore, we examined possible effects of the
inclusion of higher harmonics in the model in such cases on the
parameters of the cyclotron line and of the continuum.
Similarly to the analysis of the pulse-averaged spectra
\citep{tsy06}, the energy of the first harmonic $E_1$ was
assumed either to be a free parameter or fixed equal to the doubled
energy of the fundamental $E_0$. We have found that the
parameters of the continuum within the uncertainties do not depend on
whether the higher harmonics are included in the model. The energy and
width of the cyclotron line were found to slightly vary with
the systematic uncertainty of about 0.1 and 0.2 keV, respectively. In
the following analysis these values were taken into account
when calculating uncertainties of these parameters. Finally, to take into
account the iron fluorescent line at $\sim6.4$ keV a Gaussian line at
this energy with a fixed width 0.1 keV was included in the model as well.

The total number of spectra which were analysed in the paper exceeds
two hundreds and, therefore, it is not possible to fully discuss the results of
fitting  of the individual spectra here.
We thus present in Fig.~\ref{ex_spec} several examples of pulse-phase
resolved spectra for two luminosity levels to give some impression
on the quality of the data and typical variation of spectral parameters with pulse phase.
The best-fitting parameters are summarised in Table\,\ref{tab2}. It can be seen
from Fig.~\ref{ex_spec}  that the model adequately describes observations in a wide
energy range 3--100 keV. \citet{pot05} and \citet{nak10} have discussed
a complex structure of the V\,0332+53 spectrum near the cyclotron line. This
is also visible in the residuals of our spectra. These are likely related
to the oversimplified description of the CRSF (i.e. with a Gaussian or a Lorentzian
absorption profiles), whereas in reality it has probably a more complex shape
(as discussed e.g. by \citealt{schon07} and \citealt{mukh11}).

\section{Results}

The analysis had been carried out uniformly for all observations and a
strong variation of both the continuum and the cyclotron line
parameters with the pulse phase had been found at all luminosities.
An example of such variability for two luminosities ($L_{37}=36.9$ and
$L_{37}=12.7$) is presented in Fig.~\ref{pulphas_ex} together with the
source pulse profile in the 3--20 keV energy band. To align the phases
of individual observations we use the pulse profiles in 3--20 keV
energy range. Due to significant variation of the spin period
and the pulse profile shape with luminosity, finding a robust timing
solution coherently defining pulse phase at all luminosities is
problematic \citep{raichur10}.  Therefore, we roughly aligned pulse
profiles obtained for individual observations to have the zero phase
coincide with the main minimum of the pulse, and further on only
discuss variations of the spectral parameters over the pulse
profile. This approach does not require precise phase connection
over the whole outburst.

\begin{figure}
\includegraphics[width=0.95\columnwidth,bb=60 290 547 707, clip]{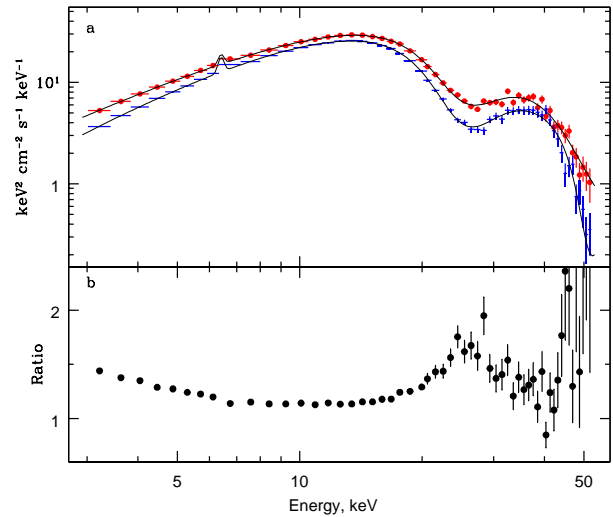}
\caption{(a) Energy spectra of V\,0332+53 obtained for the
4th (i.e. phase 0.1875-0.25; red points) and
7th bins (i.e. phase 0.375-0.4375; blue crosses) of the pulse profile (see
Fig.~\ref{pulphas_ex}) in the high luminosity state ($L_{37}=36.9$, obs. 1,
Table 1), and (b) their ratio.
The solid lines represent best-fitting models for both spectra.
\label{spectra}}
\end{figure}

As can be seen from Fig.~\ref{pulphas_ex}, the spectrum of the source is
strongly variable with pulse phase. Moreover, the pulse-phase
dependence of most parameters appears to be different at different
luminosities (see detailed descriptions in following sections).
We note that our spectral model has many parameters, some of which
can be correlated (e.g., photon index versus cutoff energy, cutoff
energy versus cyclotron line energy, see \citealt{cob02,lut11} for results and
discussion). Therefore, a change in one parameter
might in principle be compensated by a change in another parameter
with no change in the overall spectral shape. To demonstrate that this is
not generally the case, and that the spectrum does indeed significantly change its
shape, we present (as an example) in Fig.~\ref{spectra} two
spectra obtained at different pulse phases as well as their ratio. The
variability both in the continuum and near the cyclotron line energy is
apparent. Below, we discuss in details these variations.

\begin{figure*}
\hbox{
\includegraphics[width=1.05\columnwidth,bb=40 155 495 695, clip]{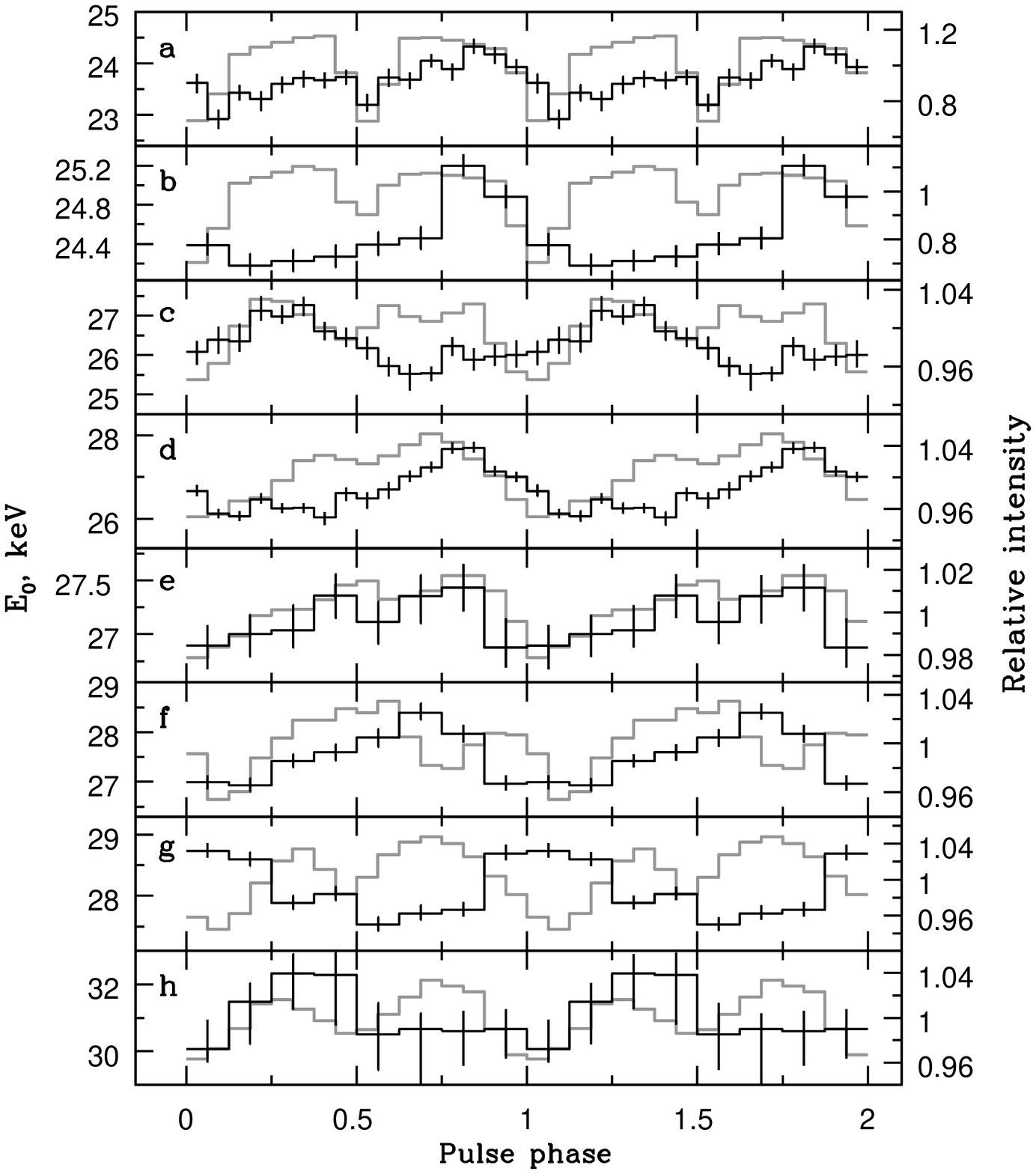}
\includegraphics[width=1.058\columnwidth,bb=61 155 520 695, clip]{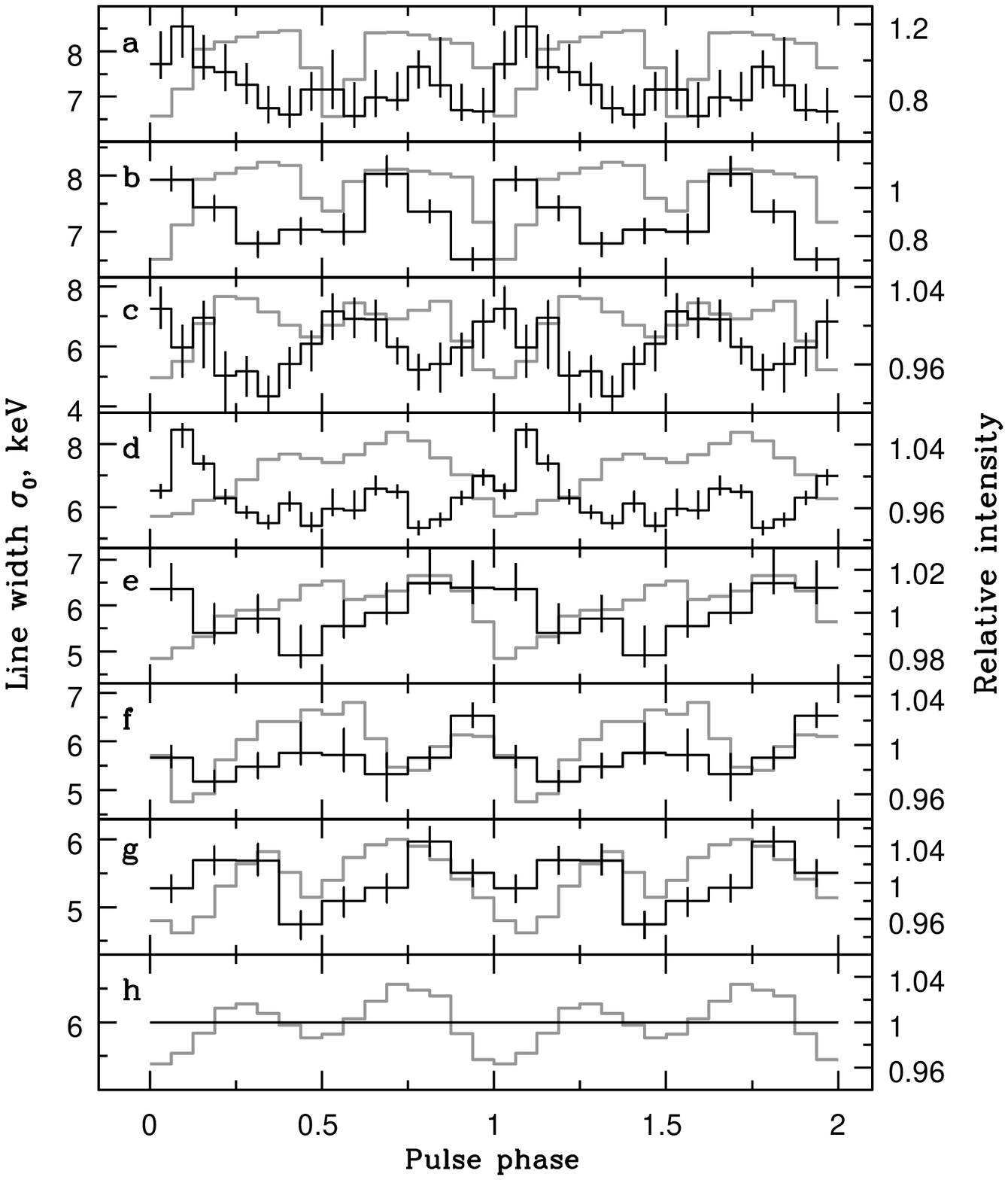}
}
\includegraphics[width=1.058\columnwidth,bb=40 155 520 695, clip]{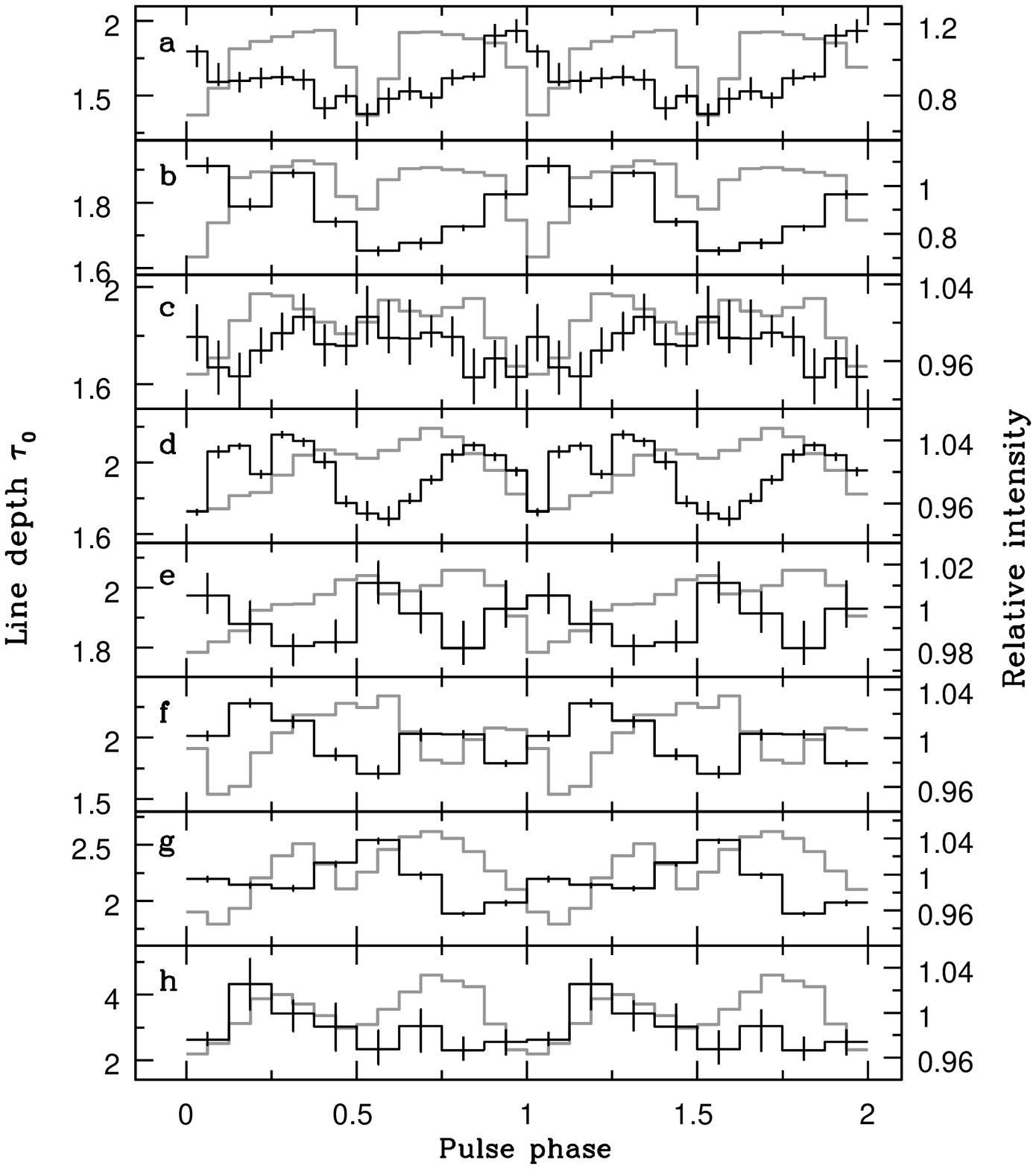}
\caption{Variations of the cyclotron line energy (top left), width (top right)
and depth (bottom) with the pulse phase for different luminosities: $L_{37}$=36.9 (a),
32.8 (b), 20.2 (c), 16.3 (d), 12.7 (e), 9.7 (f), 6.2 (g), 1.7 (h) (black histograms).
The relative columns heights according the reflection model \citep{pout13} are:
$h/R$=0.97 (a), 0.84 (b), 0.57 (c), 0.48 (d), 0.4 (e), 0.32 (f), 0.22 (g), and 0.07 (h).
Grey lines represent the corresponding normalized
pulse profiles (right axis). The line width was fixed at 6 keV for the observation with the
lowest luminosity $L_{37}=1.7$ (obs. 23).\label{ecyc}}
\end{figure*}

\subsection{Continuum}

Currently there are no commonly accepted theoretical
models to describe the spectra of accreting pulsars in details
provided by observations, so empirical models are typically used
instead. As it was noted above, the broadband spectra of V\,0332+53
(both the pulse-averaged and the phase-resolved)  were described with an
absorbed power law with exponential cutoff at high energies. The
study of the pulse-averaged spectra showed that the spectrum becomes
softer and the cutoff energy increases when luminosity drops
\citep{lut11}. Moreover, the cutoff and cyclotron line energies were
found to correlate \citep{lut11}, although the physical reasons besides this
correlation are not fully clear. In this paper, we use the same model
as it is the simplest model which seems to capture all relevant
features of the spectrum.

\subsection{Variations of the cyclotron fundamental line parameters
with phase and luminosity}

\begin{figure*}
\hbox{
\includegraphics[width=0.95\columnwidth,bb=40 155 460 675, clip]{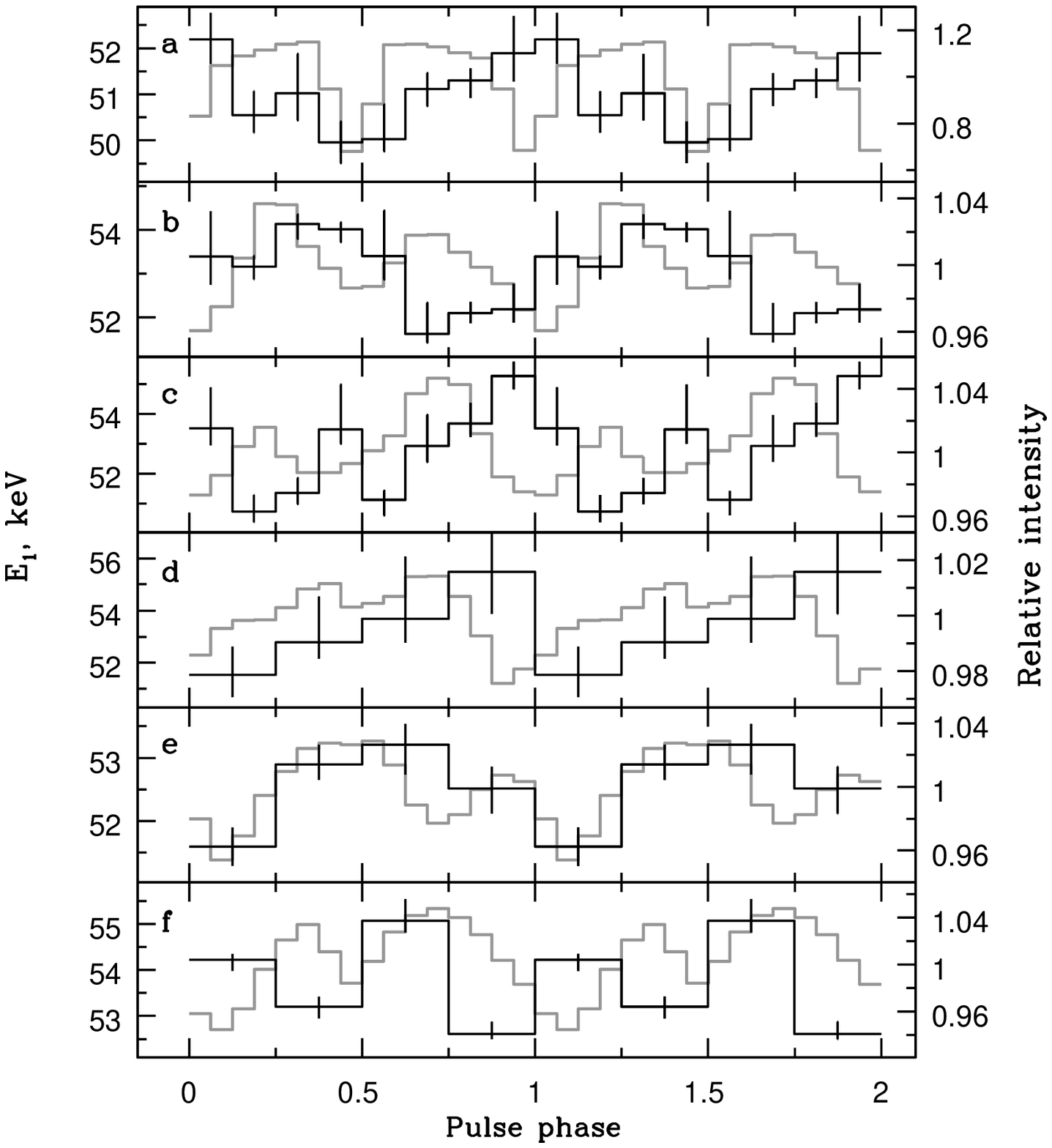}
\hspace{3mm}\includegraphics[width=1.085\columnwidth,bb=40 155 520 675, clip]{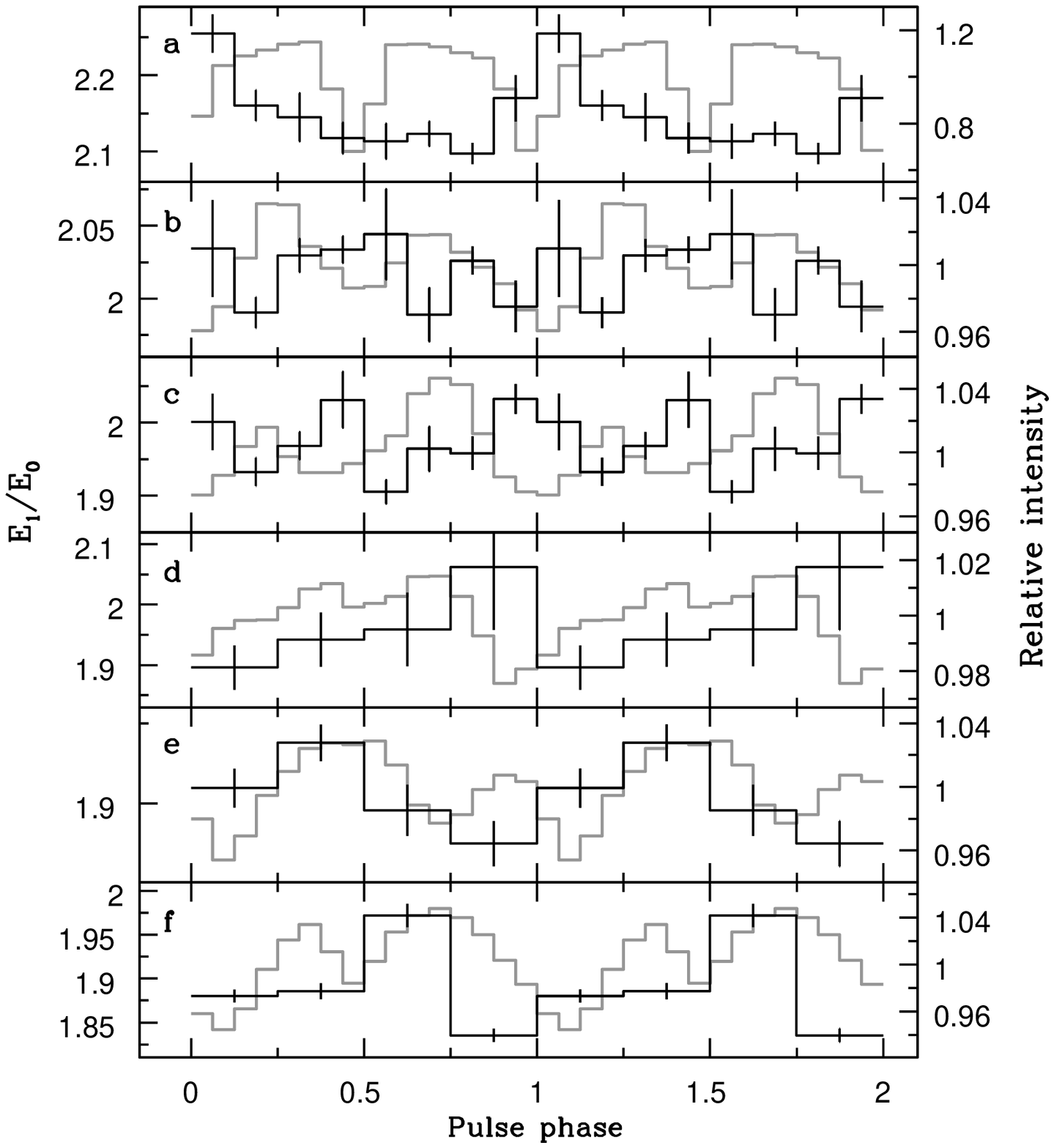}
}
\caption{Dependence of the first harmonic energy (left) and the ratio of the line
energies $E_1/E_0$ (right) on the pulse phase for different luminosities: $L_{37}$=36.9 (a),
18.6 (b), 15.3 (c), 12.7 (d), 9.7 (e), 6.2(f)  (black
histograms). Grey lines represent corresponding pulse profiles.}\label{first_harm}
\end{figure*}

As is evident from Fig.\,\ref{pulphas_ex}, the cyclotron line
parameters significantly vary with the pulse phase and the shape of
these variations (below we refer to them as `cyclotron
energy/width/depth profiles' for simplicity) depends on the source
luminosity. The observed changes of these profiles should directly
reflect changes in the structure of emission regions and their beam
patterns with luminosity. To trace this variability we constructed
the cyclotron energy/width/depth profiles for all observations from Table\,\ref{tab1}
and compared them with the corresponding pulse profiles. In
Fig.\,\ref{ecyc}, the pulse-phase dependencies of the cyclotron line
parameters are presented for eight levels of the source luminosity
$L_{37}$=36.9 (a), 32.8 (b), 20.2 (c), 16.3 (d), 12.7 (e), 9.7 (f), 6.2 (g), and 1.7 (h).

At high luminosities $L_{37}\simeq$32--37 the
cyclotron energy profile has a double-peaked shape with the maxima
roughly coinciding with the maxima of the pulse profile. With
decreasing of the luminosity to $L_{37}\simeq30$ the
profile becomes single peaked.
Such a shape remains then more or less stable down to
luminosity of $L_{37}\simeq14$. At even lower
luminosities, the maximum of the cyclotron energy profile moves to the
inter-peak minimum of the pulse profile and can be clearly seen for
luminosities below $L_{37}\simeq10$.

The evolution of the cyclotron depth profile with the luminosity
differs from the described above behaviour of the cyclotron energy
profile. In particular, practically for all luminosities its maximum
roughly corresponds to the inter-peak minimum of the pulse profile,
and only at the very low luminosities covered by observations
($L_{37}=1.7$) the maximum of the cyclotron depth
profile shifts to one of pulse profile peaks.

The cyclotron line width tends to anti-correlate with the flux
at high luminosities (Fig.\,\ref{ecyc}a, b, d) and to correlate with it
at low luminosities (Fig.\,\ref{ecyc}e, f, g). Such a behaviour
is indeed expected in the reflection model. Generally speaking, although the width
depends on several factors, it is expected to be broader in the case when we
see the illuminated neutron star surface from aside and narrower when the
illuminated surface is seen from the poles (see an extended discussion in Section 4.2).

\subsection{Variations of the energy of the first harmonic}

\citet{tsy06} showed that the centroid energy $E_1$ of the
first harmonic roughly follows the luminosity dependence of the
cyclotron line $E_0$. However, the ratio of the energies
of the first harmonic to the fundamental one does change slightly with
luminosity, from about $\sim2.2$ at highest luminosities to $\sim1.9$
at the lowest \citep{tsy06,nak10}. Taking into account that during the
pulse different parts of the emission regions are observed at
different viewing angles and that the cyclotron line and the first harmonic
could potentially be formed in slightly different regions, it would
be interesting and important to trace the behaviour of the first
harmonic energy $E_1$ and of the ratio $E_1/E_0$
both with the pulse phase and luminosity.

Note that the parameters of the first harmonic are reasonably
constrained only in a subset of all observations even for
pulse-averaged spectra \citep[see detailed discussion in][]{tsy06}. The situation is even worse for
the phase resolved analysis due to lower number of photons. Nevertheless, we
found a number of observations, covering a luminosity range from
$L_{37}\simeq5$ to $\simeq37$,
where a combination of the source intensity and exposure allowed us
to determine and constrain the parameters of the first harmonic.
Profiles of its centroid energy and corresponding ratio $E_1/E_0$ as a function of the pulse phase are presented in
Fig. \ref{first_harm}.

It is difficult to make a detailed comparison of the $E_1$
behaviour with $E_0$, however, it can be seen that their
profiles are not coincident (see Figs. \ref{ecyc} and
\ref{first_harm}). This is probably due to the fact that different parts
of emission regions are responsible for the formation of the
cyclotron line and first harmonic of the CRSF. The ratio $E_1/E_0$ is also strongly variable (up to $\sim10$\%) during the
pulse. On the average, it demonstrates also a decrease with the
luminosity, similar that found for pulse-average spectra
\citep{tsy06,nak10}.

\section{Discussion}

We now will try to describe qualitatively the observed
pulse-phase changes of the cyclotron line parameters using the recently
proposed reflection model \citep{pout13}. According to the widely
accepted paradigm \citep{bs76}, bright X-ray pulsars have relatively
tall (comparable with the neutron star radius) accretion columns at both magnetic poles.
The columns are supported by the radiative pressure of the X-rays
trapped within the optically thick accretion flow and slowly diffusing
through the sides of the columns, so the plasma is most effectively
slowed down in the core of the column, whereas the outer shell falls
almost freely \citep{ls88}. As a result, the emerging emission is
expected to be strongly beamed towards the neutron star surface due to
the scattering on fast free-falling electrons \citep{kam76,ls88}.
The formation of a cyclotron feature in the directly observed column spectrum
is rather problematic due to a strong gradient of the magnetic field along the
column and scattering on free-falling electrons \citep[see discussion in][]{nish08,nish14}.
On the other hand, such a feature can be formed in the spectrum reflected from a
neutron star atmosphere where  the magnetic field change is smaller \citep{pout13}.
The size of the spot on the neutron star surface illuminated by the accretion column
depends on the column height, which, in turn, depends
on the X-ray pulsar luminosity. The fraction of the column emission
which is intercepted by the neutron star remains, however, large at
all luminosities, so one does expect that a large fraction of the
observed emission is reflected off the surface of the neutron star.
It is not unlikely, therefore, that the observed cyclotron features
also form in the atmosphere of the neutron star. This scenario allows
to explain many important observational features of the CRSFs
that otherwise are difficult to explain. For instance, the luminosity of
V\,0332+53 in the considered outburst (and, therefore, the height of
the accretion columns) changes by factor of twenty, whereas the line
energy varies by just $\sim30$ per cent, which is hard to explain if the
line is formed within the column. On the other hand, the dipole
magnetic field over the neutron star surface decreases only by factor
of two from the magnetic pole to the equator, so smaller variations
of the CRSF energy are easier to accommodate. In fact, \cite{pout13}
showed that the expected variation of the CRSF energy in this case is
compatible with observations. Moreover, the reflection model was also capable
of explaining the observed correlation of the line energy with
luminosity. Below we will discuss qualitatively how the observed
variation of CRSF parameters with pulse phase and luminosity
can be qualitatively explained within this model.

\begin{figure}
\centering
\includegraphics[width=1.0\columnwidth,bb=80 40 515 345, clip]{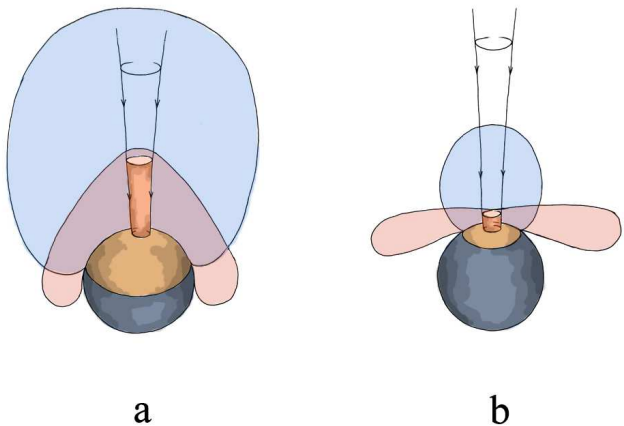}
\caption{Sketch of the angular distribution of different components for high (a) and low (b)
luminosities. Red diagram corresponds to the direct component, blue one -- to the reflected one.
Only one accretion column is drawn for clarity. \label{beaming}}
\end{figure}

\subsection{Phase dependence of the cyclotron line energy as function
of luminosity}

The quantitative comparison of results of the
phase-resolved spectroscopy with the predictions of any theoretical
models, including the reflection model, will be possible only after a
self-consistent model emerges that describes the accretion column structure and the radiation
field around the pulsar including the reflection from the neutron
star surface. Nevertheless some observational features
described above could be understood qualitatively.

\begin{figure}
\centering
\includegraphics[width=0.85\columnwidth,bb=22 277 512 674, clip]{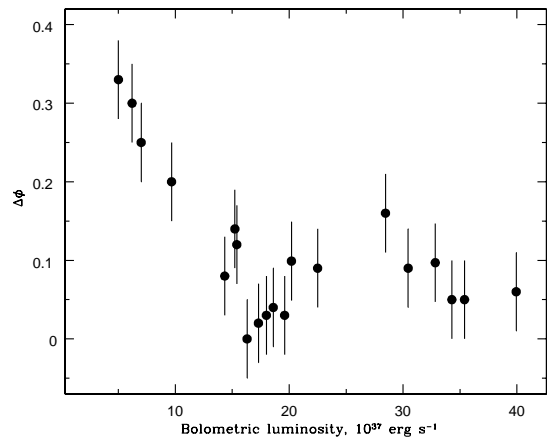}
\caption{Dependence of the phase shift between maxima of the pulse profile and the
cyclotron line energy $\Delta\phi$ on the X-ray pulsar luminosity. The error bars correspond
to phase bins. \label{corr}}
\end{figure}

If reflection off the neutron star surface is indeed responsible for
the formation of the CRSF, the observed line parameters such as its
energy, width and depth will depend on the size of illuminated
spot(s). This, in turn, depends on many factors including the
accretion columns height, beaming of the emerging radiation, the
neutron star compactness, etc. Moreover, the observable part of the
illuminated hot spots also depends on the relative position of the
observer, being a function of the pulse phase. Thus, the cyclotron
line parameters should be variable with the pulse phase.

From geometric considerations, the line energy is expected to be
maximal when one of magnetic poles is directed to the observer, so that
the polar regions with the strongest field are observed. At other angles
large fraction of reflected radiation will be scattered further from
the poles, where the field is lower, and a slight decrease in the
observed CRSF energy is expected.

It is interesting to note that the observed line energy maximum
tends to coincide with one of the total flux maxima at high luminosities
(see Figs \ref{ecyc}a-e) and it tends to be between two flux maxima
at low luminosities (Figs \ref{ecyc}f-h). A quantitative analysis this behaviour is not simple. Pulse profiles as well as
the dependence of the line energy on the phase are very different and
variable. This makes the standard correlation analysis problematic.
Instead, a phase shift between the maxima of the considered
curves can be used to quantify the potential correlation. To account for the apparent asymmetry of the profiles,
we used the centre of the segment connecting rising and declining parts
of the pulsed maximum at half of the maximum flux as a reference phase. Results of such an analysis
are presented in Fig.~\ref{corr} and they are in agreement with our expectations --
the phase shift is a relatively small ($\la 0.1$) at high luminosities and increases
at low luminosities.

In the framework of
the reflection model, the cyclotron line only contributes to the
reflected spectrum, whereas total flux at a given pulse phase includes
both reflected and directly observed emission. Therefore, the observed change in
the pulse phase behaviour of the cyclotron line energy can be
explained if the ratio between these two components
changes with luminosity. The ratio of the reflected emission to
that observed directly from the column depends on total fraction of
intercepted emission and on the beaming characteristics of the two
components. The reflected component is expected to have a broad
pencil-like beam pattern directed along the magnetic dipole axis. The
beam pattern of the emission observed directly is more complicated
and depends on luminosity.

At high luminosities, both columns are expected to contribute to the
observed flux at all pulse phases with the combined beam pattern
similar to that of the reflected component (see Fig.\,\ref{beaming}a).
Therefore, the pulse profile is expected to have two maxima
corresponding to the two illuminated spots around the two magnetic poles.
In such a situation, the maximum of the cyclotron line energy profile will
coincide with one of the pulse profile maxima, which in turn corresponds
to the pole with the smallest inclination angle to the line of sight.

\begin{figure*}
\hbox{
\hspace{2mm}
\includegraphics[width=0.95\columnwidth,bb=20 265 512 672, clip]{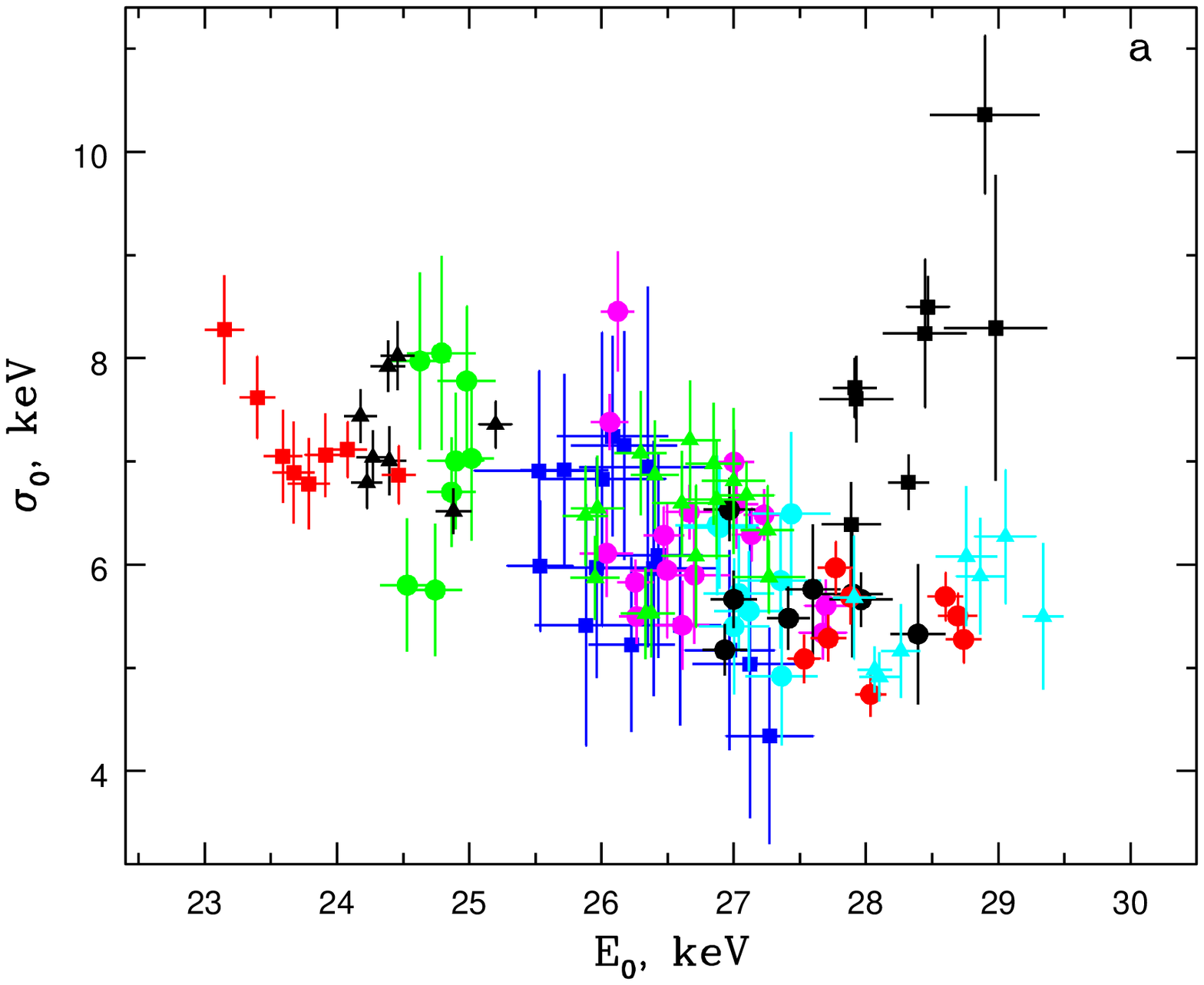}
\hspace{9mm}
\includegraphics[width=0.95\columnwidth,bb=20 265 512 672, clip]{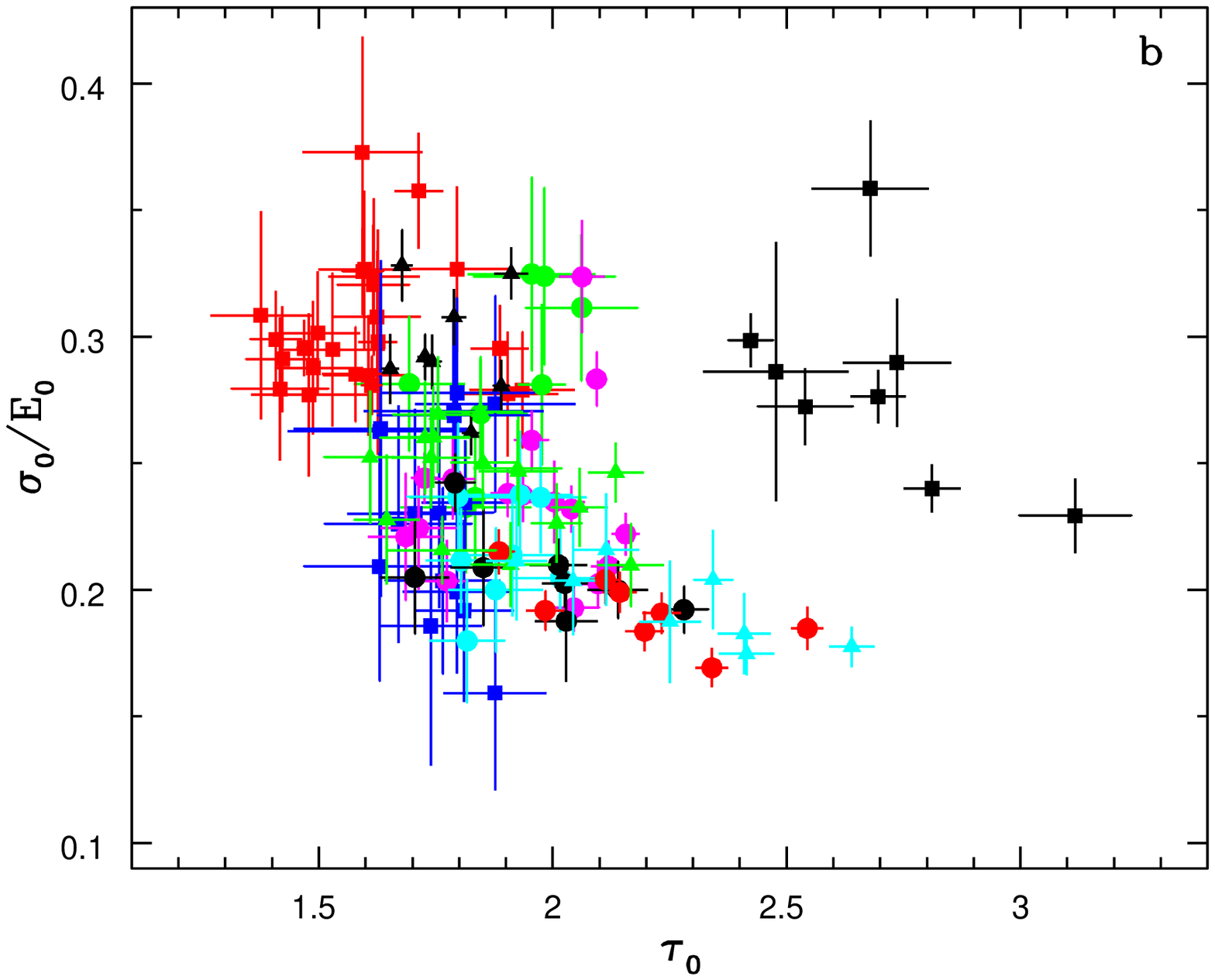}
}
\caption{(a) Cyclotron line width $\sigma_0$ versus centroid energy $E_0$.
(b) Relative cyclotron line width $\sigma_0/E_0$ versus its depth $\tau_{\rm
0}$. Different colours and symbols correspond to observations with different
luminosities (see Table \ref{tab1}): obs. 1  -- red squares,
4  -- black triangles, 5  -- green circles, 8  -- blue squares,
13 -- magenta circles, 15  -- green triangles,
17 -- cyan circles, 18  -- black circles, 19  -- cyan triangles,
20 -- red circles, 21 -- black squares.
}\label{width_vs_energy}
\end{figure*}

\begin{figure*}
\hspace{2mm}
\includegraphics[width=0.95\columnwidth,bb=30 30 520 491, clip]{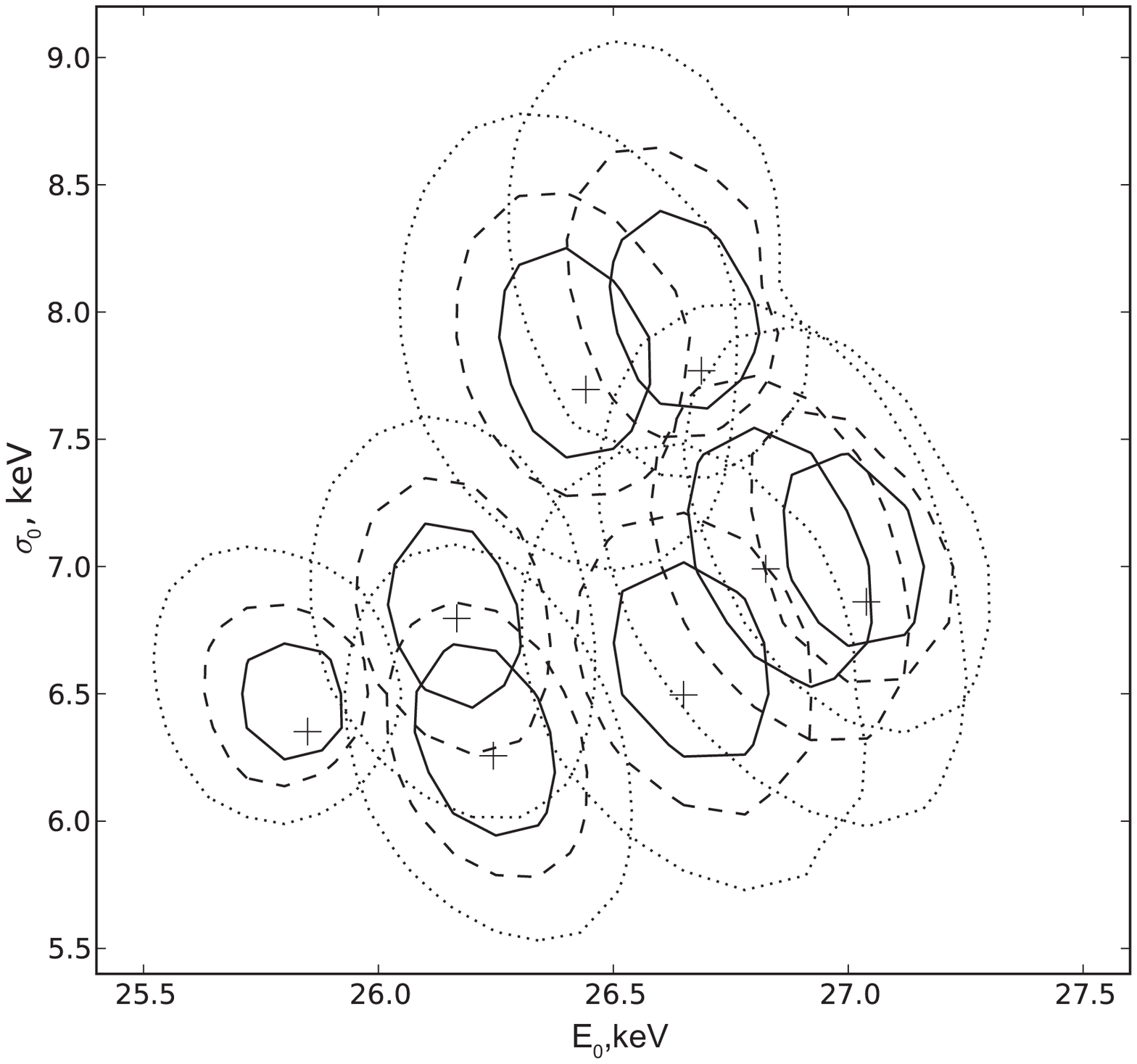}
\hspace{9mm}
\includegraphics[width=0.95\columnwidth,bb=30 28 523 491, clip]{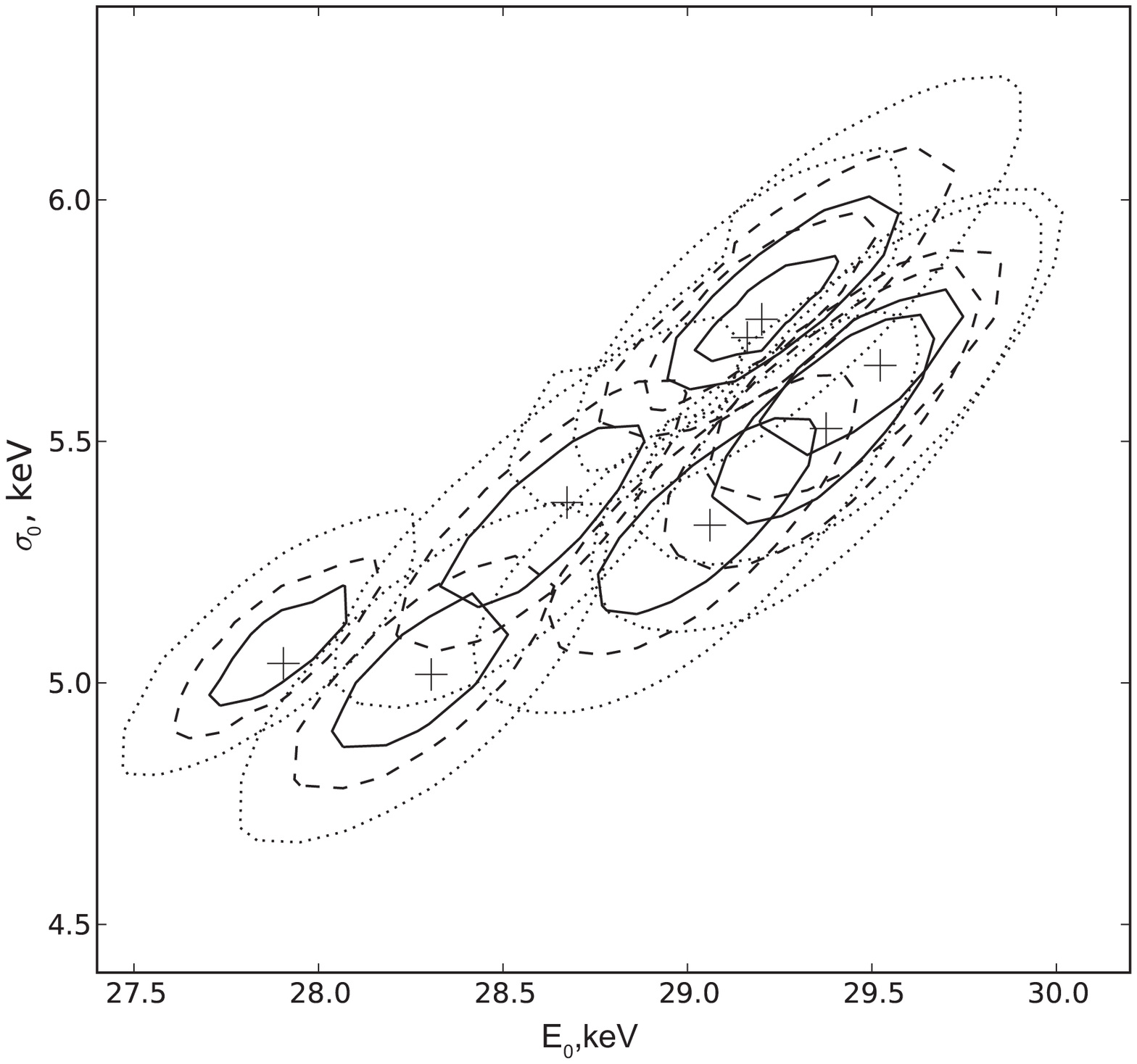}
\caption{Cyclotron line width $\sigma_0$ vs. cyclotron line energy $E_0$,
obtained using spectral fits with \protect{\sc cyclabs} (left panel) and \protect{\sc gabs}
(right panel) models for phase-resolved spectra of V\,0332+53 in observation
15 (Table \ref{tab1}).  Contours (solid, dashed, dotted lines) correspond to the
$1,2,3~\sigma$ levels, respectively.}
\label{contour}
\end{figure*}

At lower luminosities, the height of the column decreases, and the illuminated
spot around the second pole becomes less visible or even fully obscured by the neutron star.
As a result, a fan-like beam pattern from the two columns is expected
in this case to have two maxima when both columns are observed sideways (see Fig.\,\ref{beaming}).
The reflected component retains the pencil-like beaming and, therefore, is expected
to be in anti-phase with the direct emission. Even though the total fraction of the
reflected emission is expected to increase at lower luminosities,
only contribution from one pole can be observed at a time. On the
other hand, both columns are expected to be visible when the
contribution of the directly observed component is maximal, so the
overall maximum of the pulse profile shall be defined by the component
observed directly. Still, the cyclotron line is only present in the
reflected component, which is in anti-phase with the directly
observed emission. Therefore, the maximum of the cyclotron line
energy profile is expected to move away from the pulse-profile
maximum at lower luminosities in agreement with
observations.

We can find a relative column height in the framework of the
reflection model \citep{pout13}. We used equation (2) in that paper
and substituted  model parameters which
give the best fit to the observed dependence $E_{\rm cyc}$ versus $L$
(see Fig.~5  in \citealt{pout13}).
We find that the relative column height $h/R$ decreases from 0.89 for the brightest state to 0.07 for the
dimmest one (the relative column heights for all the luminosities are given in the caption to  Fig.\,\ref{ecyc}).
In framework of this model, the first described scenario with the pencil-like
reflection component domination is realised during the high-luminosity states and
relatively tall columns with $h/R > 0.5$. The second scenario, with the direct fan-like
component, is likely realised for short columns with $h/R \le 0.2$.

Finally, we note that the light bending is an important factor influencing the
visibility of illuminated spots \citep[see e.g.][]{bel02,annala10} which we did not account in out
qualitative discussion.
However, estimations of $h/R$ are done on the base of the model by
 \citet{pout13} who accounted for  light bending.

\subsection{Correlations between the parameters of the cyclotron line}

Based on the spectral analysis of several X-ray pulsars,
\citet{cob02} found that there are two types of correlations between
cyclotron line parameters: between its width $\sigma_0$ and
energy $E_0$, and between the relative cyclotron line width
$\sigma_0/E_0$ and the line depth $\tau_0$. Later,
similar correlations were reported by \citet{krey04} for pulse
phase-resolved spectra of the X-ray pulsar GX\,301--2. At the same
time no obvious correlations were found by \citet{tsy10} from the analysis of
pulse-average spectra of V\,0332+53 in a wide energy band.

\begin{table*}
\noindent
\centering
\caption{Parameters of the best fits with {\sc cyclabs} and {\sc gabs} models
to  the same pulse phase-resolved spectra of V\,0332+53 (Obs.15, $L_{37}=15.3$).}\label{tab3}
\centering
\vspace{1mm}
\small{
\begin{tabular}{c|c|c|c|c|c|c|c|c|r}
\hline\hline
Model  & $\Gamma$ & $E_{\rm cut}$ & $E_0$ & $\sigma_0$ & $\tau_0$ & $E_1$ & $\sigma_1$ & $\tau_1$ & $\chi^2$ (d.o.f.) \\
    &          & keV            & keV      & keV           &            & keV      & keV           &            &                   \\
\hline
\multicolumn{10}{c}{1st bin (phase 0.0--0.125) }  \\[0.5mm]
{\sc gabs} & $0.38\pm0.02$ & $7.49^{+0.20}_{-0.11}$ & $29.02^{+0.21}_{-0.14}$ & $5.31^{+0.18}_{-0.08}$ & $1.67^{+0.07}_{-0.03}$ & $52.2^{+0.7}_{-0.2}$ & $6.7^{+0.5}_{-0.2}$ & $2.45\pm0.11$ & 105.9(107)\\[0.5mm]
{\sc cyclabs} & $0.32\pm0.03$ & $8.36^{+0.33}_{-0.16}$ & $26.65\pm0.10$ & $6.50^{+0.36}_{-0.12}$ & $1.73\pm0.04$ & $51.6\pm0.6$ & $4.6^{+0.8}_{-0.7}$ & $4.0^{+2.3}_{-0.7}$ & 109.2(107)\\[0.5mm]
\multicolumn{10}{c}{7th bin (phase 0.75--0.875) }  \\[0.5mm]
{\sc gabs} & $0.36\pm0.04$ & $8.56\pm0.15$ & $27.90\pm0.15$ & $5.04\pm0.09$ & $1.93\pm0.07$ & $54.8\pm1.0$ & $10.5\pm0.2$ & $2.8\pm0.3$ & 108.3(107)\\[0.5mm]
{\sc cyclabs} & $0.30\pm0.04$ & $10.14\pm0.25$ & $25.85\pm0.08$ & $6.35^{+0.21}_{-0.06}$ & $2.09\pm0.04$ & $52.2\pm0.7$ & $10.5\pm1.1$ & $3.2\pm0.3$ & 107.2(107)\\[0.5mm]
\hline
\end{tabular}
}
\end{table*}

To obtain the relations $\sigma_0 - E_0$ and
($\sigma_0/E_0) - \tau_0$ we have reanalysed uniformly
all data sets from Table \ref{tab1}, dividing the corresponding pulse
profiles into eight phase bins and approximating the obtained spectra
with the model described above ({\sc cutoffpl + cyclabs}). The observations where cyclotron line
parameters were not reasonably constrained due to insufficient
statistics (effectively limiting the number of phase bins to four or less) were excluded from further analysis. Obtained results are presented
in Fig.\,\ref{width_vs_energy}. In general, its left panel is similar
to the figure 5 from the paper of \citet{tsy06}, where the dependence
of the cyclotron line width on its energy was shown for pulse phase averaged spectra. Note
a fairly strong scatter of the data points for large cyclotron energies, which is
probably caused by lower statistics at low luminosities.

Note that we used the {\sc cyclabs} model to describe the cyclotron absorption line as opposed
to many other authors  \citep[e.g.,][]{cob02,krey04,klo08}, who used
the gaussian absorption line profile  ({\sc gabs} in the {\sc xspec} package):
\begin{equation}
\exp\left[-\tau_N \exp\left(-\frac{1}{2}\left(\frac{E-E_N}{\sigma_N}\right)^2\right)\right],
\end{equation}
where $E_N$, $\sigma_N$, and $\tau_N$
are the energy, width and optical depth of the cyclotron line ($N=0$) and harmonics ($N>0$),
respectively. Both models
adequately approximate cyclotron absorption line and its harmonics (see Table\,\ref{tab3},
where best-fitting parameters for both  {\sc cyclabs}  and
{\sc gabs} models are presented). Note, that the cyclotron line energy derived from the
{\sc cyclabs} model is systematically lower (by $\sim$1--3 keV) than the
energy derived from the {\sc gabs} model \citep[see e.g.][]{mih95,nak10,tsy12}.

In this paper, we attempt not only to find spectral parameters of V\,0332+53, but also
to search for their dependence on the pulse phase and luminosity and
possible physical correlations between them. Therefore, a correlation analysis
was carried out to investigate potential intrinsic correlations of CRSF parameters for
{\sc cyclabs} and {\sc gabs} models. To illustrate obtained results, we plot in Fig.\,\ref{contour} the confidence contours
of the cyclotron line width $\sigma_0$ versus cyclotron line energy
$E_0$ obtained  with {\sc cyclabs} and {\sc gabs} models  for
luminosity $L_{37}\simeq15.3$ (obs. 15) (results for other luminosities are similar).

Several conclusions can be made from this analysis:
the line energy depends on the CRSF models (see above);
an accuracy of the line energy measurements is better for the {\sc cyclabs}
model, while the accuracy of the line width measurements is
better for the {\sc gabs} model; there is an intrinsic
correlation between $\sigma_0$ and $E_0$ for the {\sc gabs}
model, while no correlation and fairly significant scatter of the
data points or even anti-correlation are seen for the {\sc cyclabs} model.

A linear regression fit to the data shown in
both panels of Fig.\,\ref{width_vs_energy} gives a hint that there is indeed some weak anti-correlation:
$\sigma_0 = (12.7 \pm 1.6) - (0.24 \pm 0.06)  E_0$
and
$\sigma_0/E_0 = (0.30 \pm 0.03) -  (0.030 \pm 0.015)  \tau_0$.
Thus, we confirm here the previous result by \cite{tsy10} that there is no positive correlation
between the cyclotron line parameters for V\,0332+53 if the {\sc cyclabs} model is used for the cyclotron line.
Finally, we note that if  the faintest observation is excluded
from Fig.\,\ref{width_vs_energy} (black squares), the significance of the anti-correlation
is increased and in this case $\sigma_0 = (17.0 \pm 1.2) -  (0.41 \pm 0.05) E_0$ and
$\sigma_0/E_0 = (0.42 \pm 0.03) - (0.095\pm0.016)  \tau_0$.

Formation of the CRSF in the framework of the reflection model
is still pretty much work in progress, so here we restrict ourselves to qualitative discussion
of the results. The cyclotron line width $\sigma_0$ is mainly determined
by two factors. First of all, the line is thermally broadened by a factor
which depends on the plasma temperature in the line forming region and the angle between
photon momentum direction and the magnetic field. The maximum thermal line width
is realised for the radiation propagating along the magnetic field
$\Delta E_{\rm D} \approx E\, (2kT/m_{\rm e}c^2)^{1/2}$,
or about 4 keV for the X-ray pulsar
V\,0332+53 ($E \approx $~ 25 keV, $kT \approx E_{\rm cut}\sim$~5--10 keV). It is also well
known and might be important for the future quantitative analysis, that the cyclotron
line opacity strongly depends on the direction as well \citep[see e.g. Fig.\,1 in ][]{sul12}.
The second important factor is a dispersion of the magnetic field strength
over the line forming region. We focusing below on this factor.
According to the reflection model, the CRSF originates in radiation reflected
by the NS atmosphere at different latitudes. The magnetic field decreases
from poles to equator by factor of two, so the dispersion of the field within the
line forming region depends on
the size of the illuminated hot spot around the magnetic pole. If we assume that the thermal
broadening does not depend on the accretion column height, the line width will increases
with the pulsar luminosity, while the cyclotron line energy will decrease (see Fig.~4 in \citealt{pout13}).
Indeed, this
agrees with observation (see Figs\,\ref{ecyc} and \ref{width_vs_energy}a).
Strong deviations from these dependences take place only at low luminosities when the relative
column height is quite small $h/R < 0.2$ (the black squares in Fig.\,\ref{width_vs_energy}).
In this case the situation can be more complicated because the cyclotron line can be formed not only on the stellar surface but also
in the small accretion column and/or (probably hotter) plasma deceleration region.

\begin{figure}
\includegraphics[width=\columnwidth,bb=50 175 550 675,clip]{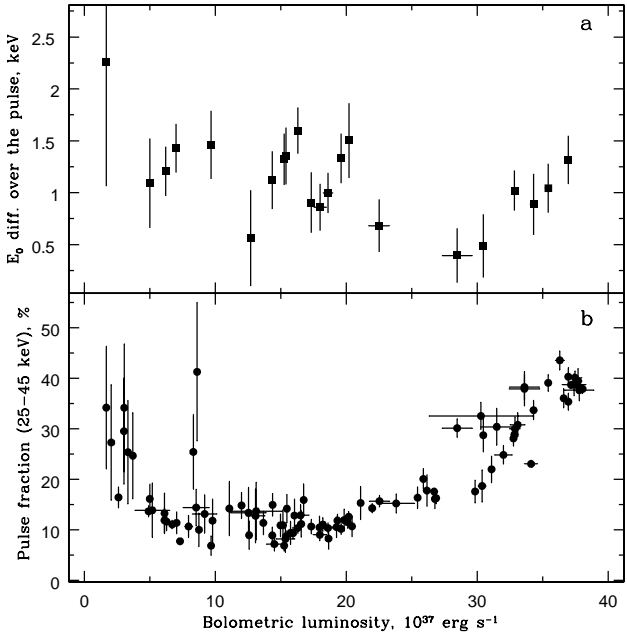}
\caption{(a) Maximal difference between cyclotron line energies over the pulse versus source luminosity.
(b) Pulse fraction in the energy range 25--45~keV versus bolometric
luminosity for V\,0332+53 (from Tsygankov et al., 2010). }\label{ecyc_diff}

\end{figure}

\subsection{Variations of the cyclotron line energy during the pulse as function of luminosity}

To character quantitatively the variability of the cyclotron line
energy with the pulse phase we introduce a parameter $\Delta E_{0,\max}$,
which is the difference between maximum and minimum values of
the cyclotron line energy $E_0$ over the pulse.
If we expect that during the pulse different parts of the emission regions are
observed, this parameter should be related to the geometrical
size of these regions and should reflect their changes with variations of
the source luminosity. In the case of super-critical sources (with the accretion
column above the surface), the parameter $\Delta E_{0,\max}$ is related to the column height and
size of the illuminated part of the neutron star surface, correspondingly
\citep{pout13}.

The dependence of $\Delta E_{0,\max}$ on the luminosity
is shown in Fig.\,\ref{ecyc_diff}a. Despite of some scattering, we see
that $\Delta E_{0,\max}$ decreases with increase of the
luminosity up to $L_{37}\approx 30$. At higher
luminosities $\Delta E_{0,\max}$ increases quickly practically up
to values, typical for low luminosities. To quantify this,
we approximated the $\Delta E_{0,\max} - L_{37}$ dependence
with the linear fit in different luminosity ranges. We found that
$\Delta E_{0,\max} =  (1.65\pm 0.20) - (0.03\pm0.01) L_{37}$ for
$L_{37}\leq 30$ and $\Delta E_{0,\max} = (-2.4 \pm 1.8) + (0.10\pm0.05)  L_{37}$ for
$L_{37}\geq 30$. This gives $>99$ per cent significance that the
difference in slopes is real. Finally, we note that although it is hard to interpret such behaviour
of $\Delta E_{0,\max}$, it is similar to the behaviour of the pulse
fraction with the source luminosity (Fig.\,\ref{ecyc_diff}b).\\

\section{Summary}

The pulse phase-resolved spectroscopy of X-ray pulsars provides a lot
of additional information in a comparison with the analysis of the
pulse-averaged spectra, therefore it can be used as a powerful
instrument for the discrimination between different models for the
X-ray pulsars emission mechanisms. Up to date the pulse
phase-resolved spectroscopy has been done for several sources.
However, until now the limitations of observational data did not allow to investigate
 the influence of the source luminosity which
determines geometrical structure of the emitting region.

In this paper we presented results of the pulse phase-resolved
spectroscopy of the transient X-ray pulsar V\,0332+53 based on the
{\it RXTE} observations carried out during the powerful outburst in Dec
2004 -- Feb 2005. For the first time such a study was performed for
the same source in a wide luminosity range $L_{37}\simeq(1-40)$.

It was shown that both continuum and cyclotron line parameters are
strongly variable with the pulse phase at all luminosities. The most
important conclusion is that all spectral parameters are variable not
only with the pulse phase, but the pattern of this variability with
respect to the pulse profile is different for different luminosities.
Such a behaviour reflects directly changes in the structure of
emission regions and their beam pattern with the luminosity changes.

For instance, at very high luminosities
$L_{37}\simeq(32-37)$, the cyclotron energy
profile has double-peaked shape almost coinciding
with the pulse profile. With the decrease of the source intensity to
$L_{37}\simeq30$ the cyclotron energy profile
is transformed to the single broad peak with the maximum at the same phases where
the pulse profile peaks. Such a shape remains more or less
stable till the luminosity of $L_{37}\simeq14$. At even lower luminosities below
$L_{37}\simeq10$ the maximum of the
cyclotron energy profile moves to the inter-peak minimum of the
pulse profile.

We interpreted these observational results qualitatively in the
framework of the recently proposed reflection model for the cyclotron
line formation \citep{pout13}. Nevertheless, it is necessary to say
that deep observations with more sensitive instruments in hard X-rays
(e.g., {\it NuSTAR}) are needed to make quantitative conclusions and verify
an applicability of competitive models \citep[see, e.g.,][]{nish14}.

\section*{Acknowledgements}

This work was supported by the Russian Scientific Foundation grant 14-12-01287 (AAL, AAM).
Authors also acknowledge support for grant RFBR 12-02-97006-r-povolzhe-a,
Deutsche Forschungsgemeinschaft (DFG)  grant WE 1312/48-1,
the COST Action MP1304 and the Magnus Ehrnrooth Foundation (VFS),
the Deutsches Zentrums f\"ur Luft-und Raumfahrt and DFG  grant DLR~50~OR~0702 (VD),
the  Saint Petersburg State University grants 6.0.22.2010, 6.38.669.2013, 6.38.18.2014 (DIN),
and the Academy of Finland grant 268740 (JP). The research used the data obtained from
the HEASARC Online Service provided by the NASA/GSFC.

\bibliographystyle{mn2e}

\label{lastpage}

\end{document}